\documentclass[twocolumn]{aastex63}
\usepackage{multirow}
\usepackage{amsmath}
\usepackage[figuresright]{rotating}
\usepackage{threeparttable}
\usepackage{subfigure}
\usepackage{epic,eepic}
\usepackage{graphicx}
\usepackage{longtable}
\usepackage{float}
\usepackage{lineno}
\usepackage{pifont}
\usepackage{hyperref}
\usepackage{gensymb}
 \usepackage{textcomp, gensymb}

\shorttitle{\rm Activity and habitability of M stars}
\shortauthors{Li et al.}

\begin{document}

\title{\rm Ultraviolet and Chromospheric activity and Habitability of M stars}
 
\correspondingauthor{Song Wang}
\email{songw@bao.ac.cn}

\author{Xue Li}
\affiliation{Key Laboratory of Optical Astronomy, National Astronomical Observatories, Chinese Academy of Sciences, Beijing 100101, China}
\affiliation{School of Astronomy and Space Sciences, University of Chinese Academy of Sciences, Beijing 100049, China}

\author{Song Wang}
\affiliation{Key Laboratory of Optical Astronomy, National Astronomical Observatories, Chinese Academy of Sciences, Beijing 100101, China}
\affiliation{Institute for Frontiers in Astronomy and Astrophysics, Beijing Normal University, Beijing 102206, China}

\author{Henggeng Han}
\affiliation{Key Laboratory of Optical Astronomy, National Astronomical Observatories, Chinese Academy of Sciences, Beijing 100101, China}

\author{Huiqin Yang}
\affiliation{Key Laboratory of Optical Astronomy, National Astronomical Observatories, Chinese Academy of Sciences, Beijing 100101, China}
\affiliation{Institute for Frontiers in Astronomy and Astrophysics, Beijing Normal University, Beijing 102206, China}

\author{Chuanjie Zheng}
\affiliation{Key Laboratory of Optical Astronomy, National Astronomical Observatories, Chinese Academy of Sciences, Beijing 100101, China}
\affiliation{School of Astronomy and Space Sciences, University of Chinese Academy of Sciences, Beijing 100049, China}

\author{Yang Huang}
\affiliation{School of Astronomy and Space Sciences, University of Chinese Academy of Sciences, Beijing 100049, China}
\affiliation{Key Laboratory of Optical Astronomy, National Astronomical Observatories, Chinese Academy of Sciences, Beijing 100101, China}

\author{Jifeng Liu}
\affiliation{Key Laboratory of Optical Astronomy, National Astronomical Observatories, Chinese Academy of Sciences, Beijing 100101, China}
\affiliation{School of Astronomy and Space Sciences, University of Chinese Academy of Sciences, Beijing 100049, China}
\affiliation{Institute for Frontiers in Astronomy and Astrophysics, Beijing Normal University, Beijing 102206, China}
\affiliation{New Cornerstone Science Laboratory, National Astronomical Observatories, Chinese Academy of Sciences, Beijing 100012, People's Republic of China}

\let\cleardoublepage\clearpage
\begin{abstract}

M-type stars are crucial for stellar activity studies since they cover two types of magnetic dynamos and particularly intriguing for habitability studies due to their abundance and long lifespans during the main-sequence stage.
In this paper, we used the LAMOST DR9 catalog and the GALEX UV archive data to investigate the chromospheric and UV activities of M-type stars.
All the chromospheric and UV activity indices clearly show the saturated and unsaturated regimes and the well-known activity-rotation relation, consistent with previous studies.
Both the FUV and NUV activity indices exhibit a single-peaked distribution, while the {\rm H$\alpha$} and \rm {Ca \scriptsize{\uppercase\expandafter{\romannumeral2}} \normalsize H$\&$K} indices show a distinct double-peaked distribution.
The gap between these peaks suggests a rapid transition from a saturated population to an unsaturated one.
The smoothly varying distributions of different subtypes suggest a rotation-dependent dynamo for both early-type (partly convective) to late-type (fully convective) M stars.
We identified a group of stars with high UV activity above the saturation regime (log$R^{\prime}_{\rm NUV} > -2.5$) but low chromospheric activity, and the underlying reason is unknown.
By calculating the continuously habitable zone and the UV habitable zone for each star, we found about 70\% stars in the total sample and 40\% stars within 100 pc are located in the overlapping region of these two habitable zones, indicating a number of M stars are potentially habitable.
Finally, we examined the possibility of UV activity studies of M stars using the China Space Station Telescope.

\end{abstract}


\section{INTRODUCTION}
\label{intro.sec}

M-type stars are thought to exhibit stronger magnetic activity compared to other types of stars. 
Numerous studies have been conducted to investigate the stellar activities of M-type stars, by using X-ray \citep[e.g.,][]{2013MNRAS.431.2063S,2016Natur.535..526W}, H$\alpha$ emission \citep[e.g.,][]{2014ApJ...795..161D,2017ApJ...834...85N},  Ca \scriptsize{\uppercase\expandafter{\romannumeral2}} \normalsize H$\&$K emission \citep[e.g.,][]{2017A&A...600A..13A,2018A&A...616A.108B,2020NatAs...4..658L}, UV emission \citep[e.g.,][]{2013MNRAS.431.2063S,2018AJ....155..122S,2023arXiv230506561R}, and optical flare \citep[e.g.,][]{2017ApJ...849...36Y}, etc.
These investigations aim to understand the manifestations of stellar activity and its connection to the stellar dynamo.
For early M stars, they typically follow the solar-type dynamo mechanism ($\alpha$-$\Omega$ dynamo or tachocline dynamo). 
The generation of magnetic fields occurs in their deep convection zones due to the interior radial differential rotation, and the magnetic fields are then amplified through the interaction between magnetic flux tubes and convection processes \citep[e.g.,][]{1975ApJ...198..205P,2000nlod.book.....R}.
On the other hand, late M stars, which are fully convective, lack a tachocline and exhibit a different dynamo mechanism, such as the $\alpha^2$ dynamo.
Therefore, M-type stars offer a unique opportunity to study two different magnetic dynamos within a single stellar type.

Since M-type stars constitute approximately 70\% of the total stellar population in the Milky Way \citep{2009AIPC.1094..977B}, there is significant interest in investigating the habitable zones and potentially habitable planets around M stars.
Thanks to recent space missions like Kepler and TESS, several habitable planets have been identified orbiting M stars, such as Trappist-1 d-g, Proxima Cen b, K2-18 b, etc. \citep{2023MNRAS.522.1411S}.
However, there is an ongoing debate about the habitability of the surroundings around M-type stars \citep[e.g.,][]{2018AJ....155..122S, 2023arXiv230506561R, 2023MNRAS.522.1411S}.
Most previous studies have focused on a quite limited sample, and a large sample with accurate UV emission measurements may provide further insights into this question.

The Large Sky Area Multi-Object Fiber Spectroscopic Telescope (hereafter LAMOST, also called the GuoShouJing Telescope), is an innovative telescope designed with both a large-aperture and a wide field of view for astronomical spectroscopic survey.
The unique design of LAMOST enables it to take more than 3000 spectra in a single exposure to a limiting magnitude as faint as $r =$ 19 mag at the low-resolution \citep{2012RAA....12.1197C}. 
The low-resolution spectroscopic survey began in October 2011, with a wavelength coverage of 3690--9100\AA\ and a resolution of $R \sim$ 1800. 
As of 2021 June, LAMOST Data Release 9 (DR9) published 11,226,252 low-resolution spectra, including 832,755 M Giants, Dwarfs and Subdwarfs\footnote{http://www.lamost.org/dr9/v1.0/}.
The vast sources will contribute significantly to our studies of M-type stars.

The Galaxy Evolution Explorer (hereafter GALEX) is a NASA Small Explorer mission designed to conduct an all-sky survey in the ultraviolet (UV) band \citep{2007ApJS..173..682M}.
It has observed in the far-UV (FUV, $\lambda_{\rm eff}\sim1528\AA$, 1344--1786\AA) and near-UV (NUV, $\lambda_{\rm eff}\sim2310\AA$, 1771--2831\AA) bands. 
The latest catalog GR6$+$7 \citep{2017ApJS..230...24B}, released by GALEX in June 2017, includes observations from an All-Sky Imaging Survey (AIS, $t_{\rm exp} =$ 100 secs) and a Medium-depth Imaging Survey (MIS, $t_{\rm exp} =$ 1500 secs), for a total of 82,992,086 objects. 
The detection limit is $\approx$20 mag in the FUV band and $\approx$21 mag in the NUV band for AIS, and $\approx$22.7 mag in both the FUV and NUV bands for MIS \citep{2017ApJS..230...24B}. 

In this work, we studied stellar UV activity with GALEX data and chromospheric activity with LAMOST DR9 low-resolution data.
In section \ref{sample.sec}, we introduce the sample construction and the calculation of atmospheric parameters.
Section \ref{activity.sec} describes in detail the calculation of stellar activities and rotation periods, the distributions of different activity indices, and the activity-rotation relation.
Section \ref{flare.sec} presents UV flares detected in our sample. 
We discuss the habitability of the sample in section \ref{hz.sec}.
In section \ref{discussion.sec}, we discuss the possibility of using CSST data to study stellar activity and habitable zones in the future. 
We summarize our study in section \ref{sum.sec}.

\section{Sample of M-type stars}
\label{sample.sec}

\subsection{Sample construction}
\label{lamost.sec}

LAMOST DR9 low-resolution data released atmospheric parameters for 832,755 spectra from 588,276 M-type stars, by fitting the spectra to BT-Settl atmospheric models \citep{2021RAA....21..202D}.
The spectra with $S/N_{\rm r}>10$, $S/N_{\rm i}>10$ and $S/N_{\rm g}>7$ were selected.
We cross-matched the M-star catalog and the LAMOST LRS Stellar Parameter Catalog of A, F, G and K Stars, and remove the common sources from our sample.
As a result, there are 237,942 M stars in our initial sample.

The GALEX $PhotoObjAll$ catalog \citep{2017ApJS..230...24B} contains FUV and NUV photometric data for 292,296,119 sources observed by both the AIS and MIS.
We performed a cross-match between the M-star sample from LAMOST and the GALEX $PhotoObjAll$ catalog using a match radius of 3$\arcsec$ via the CasJobs\footnote{https://galex.stsci.edu/casjobs/}. 
The closest neighbor within the radius was considered the true counterpart.
Sources with flags of ``nuv\_artifact" $>$ 1 or ``fuv\_artifact" $>$ 1 were excluded.
This resulted in 15,952 M stars with available FUV or NUV magnitudes. 
All the {\it GALEX} data used in this paper can be found in MAST: \dataset[10.17909/T9H59D]{http://dx.doi.org/10.17909/T9H59D} and \dataset[10.17909/T9CC7G]{http://dx.doi.org/10.17909/T9CC7G}.

Gaia eDR3 provided distance measurements for approximately 1.47 billion objects \citep{2021AJ....161..147B}.
We cross-matched the M-star sample (with UV photometry) with Gaia eDR3 distance catalog using a match radius of 3$\arcsec$. 
In order to have accurate distance estimations, we excluded the objects with distances larger than 5 kpc and relative parallax uncertainties larger than 0.2.

During this process, we found that spatially close sources in Gaia catalog may be mistakenly identified as one source by GALEX due to the low resolution ($R =$ 1.5$\arcsec$/pixel; FWHM$_{\rm FUV}$ $\sim$ 4.2$\arcsec$; FWHM$_{\rm NUV}$ $\sim$ 5.3$\arcsec$)\footnote{https://archive.stsci.edu/missions-and-data/galex}.
We therefore searched for sources with multiple counterparts within 10 arcsecs in the Gaia eDR3 catalog. We removed those sources when the luminosity ratio between the brightest and faintest  counterparts was less than 100 in any of the $G$, $BP$ or $RP$ bands.
This step yielded a sample of 14,119 M stars with atmospheric parameters from LAMOST, UV photometry from GALEX and distance measurements from Gaia.

\subsection{Sample cleaning}
\label{selection.sec}

The M sample suffers from contamination by binaries, pulsating variables, young stellar objects (YSOs), white dwarfs, galaxies, and active galactic nucleus (AGNs), etc.
We employed a series of methods to clean our sample.

\subsubsection{binaries and pulsating variables}
\label{binaries.sec}

First, we searched for binaries and pulsating variables using the light curves.
We conducted a list of variable stars from various surveys, including 
Kepler \citep[e.g.,][]{2013MNRAS.432.1203M,2014ApJS..211...24M, 2016AJ....151...68K,2019ApJS..244...21S}, 
K2 \citep{2020A&A...635A..43R}, 
ZTF \citep{2020ApJS..249...18C}, 
ASAS-SN \citep[e.g.,][]{2018MNRAS.477.3145J, 2023MNRAS.519.5271C}, 
Catalina \citep{2014ApJS..213....9D}, 
WISE \citep{2018ApJS..237...28C}, 
Gaia \citep{2023A&A...674A..14R}, 
TESS \citep[e.g.,][]{2022RNAAS...6...96H, 2022ApJS..258...16P}, 
GCVS \citep{2017ARep...61...80S}, 
OGLE \citep{2016AcA....66..405S},
LAMOST DR7 \citep{2022A&A...660A..38W},
MEarth \citep{2016ApJ...821...93N},
and HATNet \citep{2011AJ....141..166H}. 
We cross-matched our sample with these variable catalogs using a matching radius of 3$\arcsec$. 
For sources observed by multiple surveys, we selected the variable type and period based on the priority sequence as mentioned above.
The eclipsing binaries and pulsating variables were removed from our sample.
For sources that are not found in these catalogs, we downloaded TESS light curves using the {\it Lightkurve} package \citep{2018ascl.soft12013L} and employed the Lomb-Scargle method \citep{1976Ap&SS..39..447L,1982ApJ...263..835S,2018ApJS..236...16V} to estimate the periods.
Through visual check of the folded light curves, we identified and threw away the eclipsing binaries (EA and EB) and possible pulsating variables.

Second, we identified spectroscopic binaries by calculating the radial velocity (RV) variation using the LAMOST DR9 low-resolution spectra (LRS, $R \sim$ 1800) and medium-resolution spectra (MRS, $R \sim$ 7500) catalogs.
A source was considered as a binary and removed if the RV variation is larger than 10 km/s.
In addition, we downloaded the LAMOST DR9 medium-resolution spectra of the M stars, 
and estimated RVs of spectra with the cross correlation function maximization method.
We removed spectroscopic binaries by selecting cross correlation functions with double peaks.
We also selected spectroscopic binaries or multiples from previous catalogs which aim at detecting multiline spectroscopic systems \citep{2021ApJS..256...31L,2022ApJS..258...26Z}.

Third, a number of works have tried to detect white dwarfs, white dwarf–main-sequence (WDMS) binaries, and binaries containing two main-sequence stars, using the GALEX data \citep{2011MNRAS.411.2770B}, LAMOST spectra \citep{2013AJ....146...82R, 2018MNRAS.477.4641R, 2020ApJ...905...38R}, Gaia photometry \citep{2018MNRAS.480.4505J, 2021MNRAS.508.3877G} and astrometry \citep{2021MNRAS.506.2269E}.
We excluded sources that appeared in these catalogs.
\cite{2021MNRAS.506.5201R} defined an area of WDMS binaries in the Hertzsprung–Russell diagram using the Gaia eDR3 magnitudes, and we applied this area to identify and exclude WDMS binary candidates from our sample.

\subsubsection{young stellar objects}
\label{yso.sec}

YSOs are a main source of contamination in our M-star sample.
We cross-matched our sample with these previous catalogs \citep{2016MNRAS.458.3479M,2018A&A...619A.106G,2019MNRAS.487.2522M,2021ApJS..254...33K,2023A&A...674A..21M,2023A&A...674A..14R} to select YSO candidates.
\cite{2016MNRAS.458.3479M} presented a catalog of Class I/II and III YSO candidates by using the 2MASS and WISE photometric data, together with Planck dust opacity values. 
By adding the {\it Gaia} database, \cite{2019MNRAS.487.2522M,2023A&A...674A..21M} presented new catalogs of YSOs. 
The YSOs identified in \cite{2021ApJS..254...33K} were selected based on MIR observations of {\it Spitzer}, while the YSO catalog in \cite{2018A&A...619A.106G} were chosen based on observations from the ESO–VISTA NIR survey. 
Furthermore, \cite{2020ApJ...902..114W} classified a source as YSO candidate if the Planck dust opacity value is higher than $1.3\times 10^{-5}$, and if the absolute magnitudes satisfy the conditions $J-H<1-(H-K_{s})$ or $W1-W2>0.04$.
We applied the same criterion to clean our sample.

\subsubsection{other type objects}
\label{simbad.sec}

Gaia DR3 classified 12.4 million sources into 9 million variable stars (22 variability types), thousands of supernova explosions in distant galaxies, 1 million active galactic nuclei, and almost 2.5 million galaxies \citep{2023A&A...674A..14R}. 
We only kept the objects classified as `SOLAR LIKE' (i.e., with rotation signals or flares), `RS' (i.e., possible RS Canum Venaticorum variable), or `LPV' (i.e., with long period signals) types \citep[see][for more details]{2023A&A...674A..14R}.
In addition, sources with $(G_{\rm bp}-G_{\rm rp})_{0} < 1.25$ mag \citep{2019ApJS..244....8Z} were excluded as they were misidentified M-type stars.
Finally, we cross-matched with the SIMBAD database \citep{2000A&AS..143....9W}, and removed the sources classified as ``AGN", ``Galaxy", ``white dwarf", ``RR lyrae", ``RS CVn", ``Gamma Dor", ``spectroscopic binary", ``eclipsing binary", ``multiple object", ``Orion*", ``TTau", etc.

\subsubsection{objects with low-quality spectra}
\label{bad_spec.sec}

Finally, we downloaded LAMOST DR9 low-resolution spectra of the sample stars and have a visual check of the spectra.
Those spectra of poor quality (e.g., too many masks, negative fluxes) were removed.

\begin{figure}[htp]
   \center
   \includegraphics[width=0.5\textwidth]{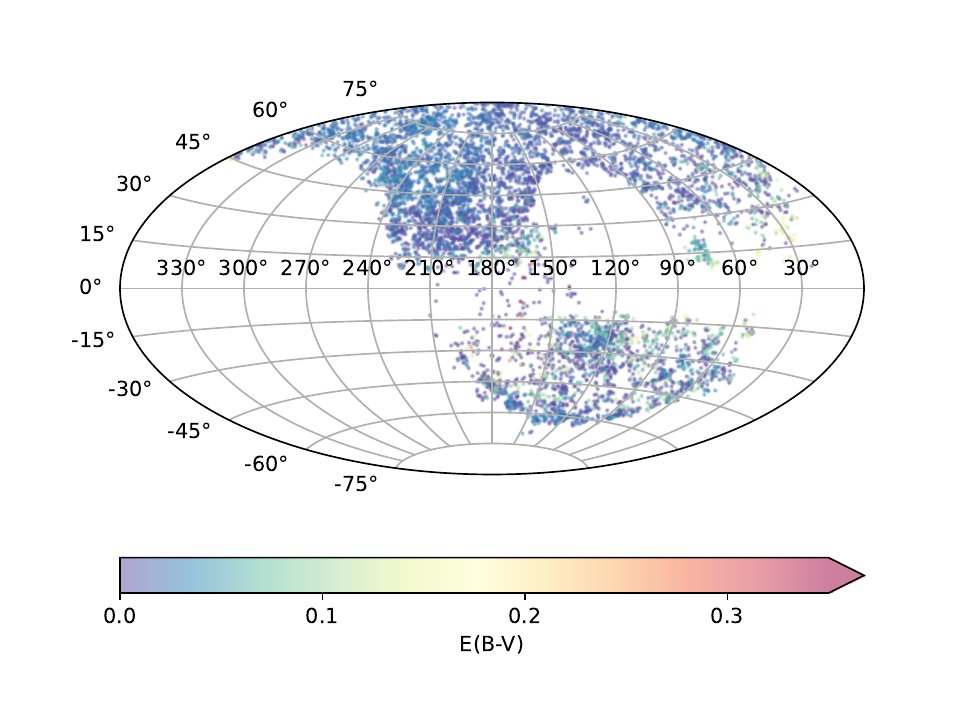}
   \includegraphics[width=0.5\textwidth]{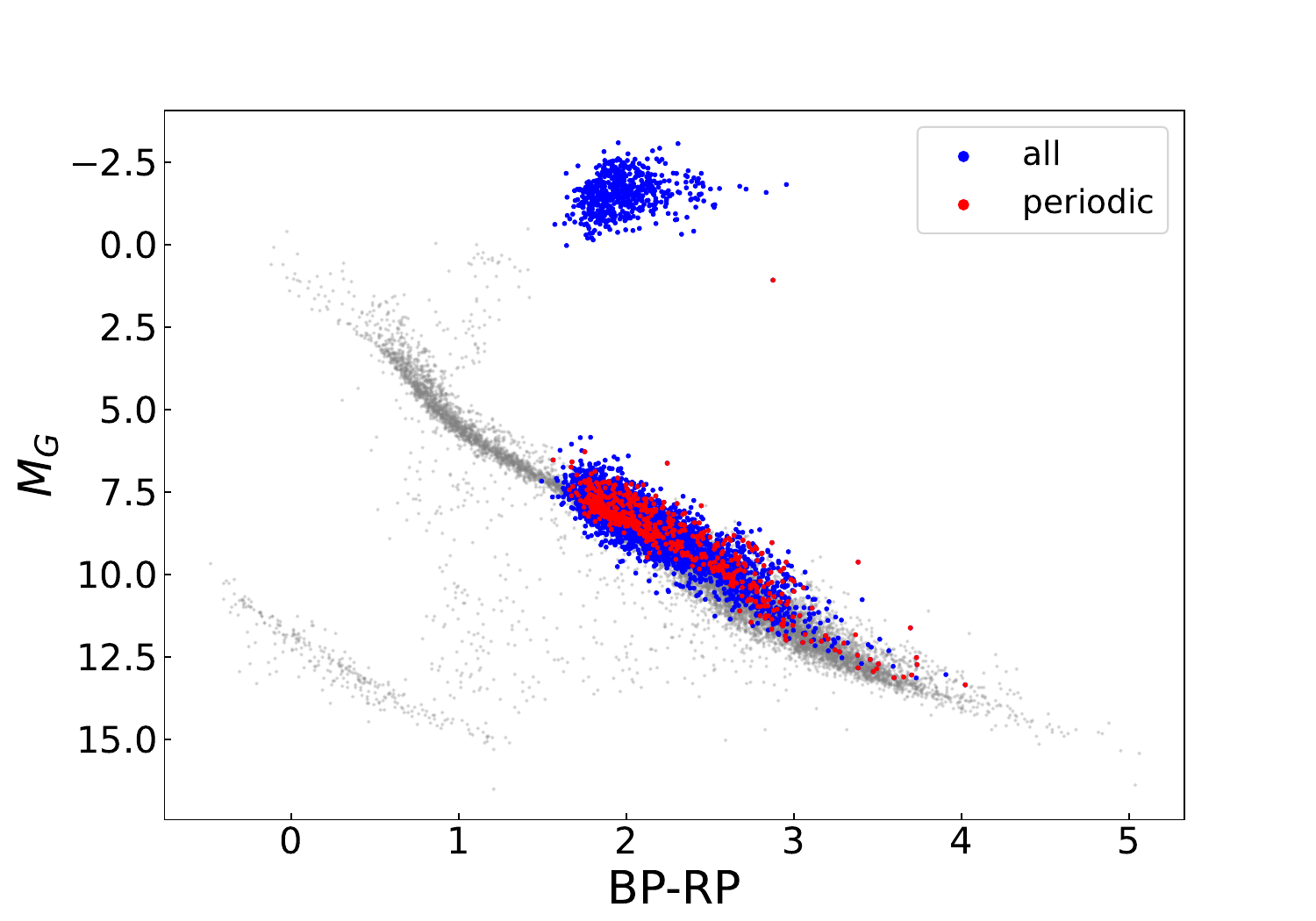}
   \caption{Top panel: Galactic distribution of the sample stars. The color bar means the extinction value. Bottom panel: Hertzsprung–Russell diagram of all sample. Blue points means all sample, red points means the sample with period. The grey points are the stars from {\it Gaia} eDR3 with distances $d <$ 200 pc, $G_{\rm mag}$ between 4--18 mag, and Galactic latitude |$b$| $>$ 10. No extinction was corrected for these stars.}
   \label{galatic.fig}
\end{figure}

In summary, we obtained a total of 6,629 stars, including 5907 dwarfs and 722 giants.
Figure \ref{galatic.fig} shows the sky map of all objects in galactic coordinates, and the positions of these targets in the Hertzsprung–Russell diagram. 
In this paper, only dwarfs were studied and discussed in detail, and the activity of giants were briefly described in the appendix \ref{giants.sec}.

\subsection{Atmospheric parameters}

For objects with one observation, we used the atmospheric parameters from the corresponding spectrum.
For objects with multiple observations, the atmospheric parameters and their uncertainties were derived following \citep{2020ApJS..251...15Z},
\begin{equation} \label{eqweight}
\overline{P} = \frac{\sum_k w_k \cdot P_{k}}{\sum_k w_k}
\end{equation}
and
\begin{equation}
\sigma_w(\overline{P}) = \sqrt{\frac{N}{N-1}\frac{\sum_k w_k \cdot (P_{k} - \overline{P})^2}{\sum_k w_k}}.
\end{equation}
The index $k$ is the epoch of the measurements of parameter $P$ (i.e., $T_{\rm eff}$, log$g$, and [Fe/H]) for each star, and the weight $w_k$ is estimated of the square of the {\it S/N} for each spectrum.

\begin{figure*}[htp]
\centering
    \subfigure[]{
    \label{teff_all_p.fig}
    \includegraphics[width=0.49\textwidth]{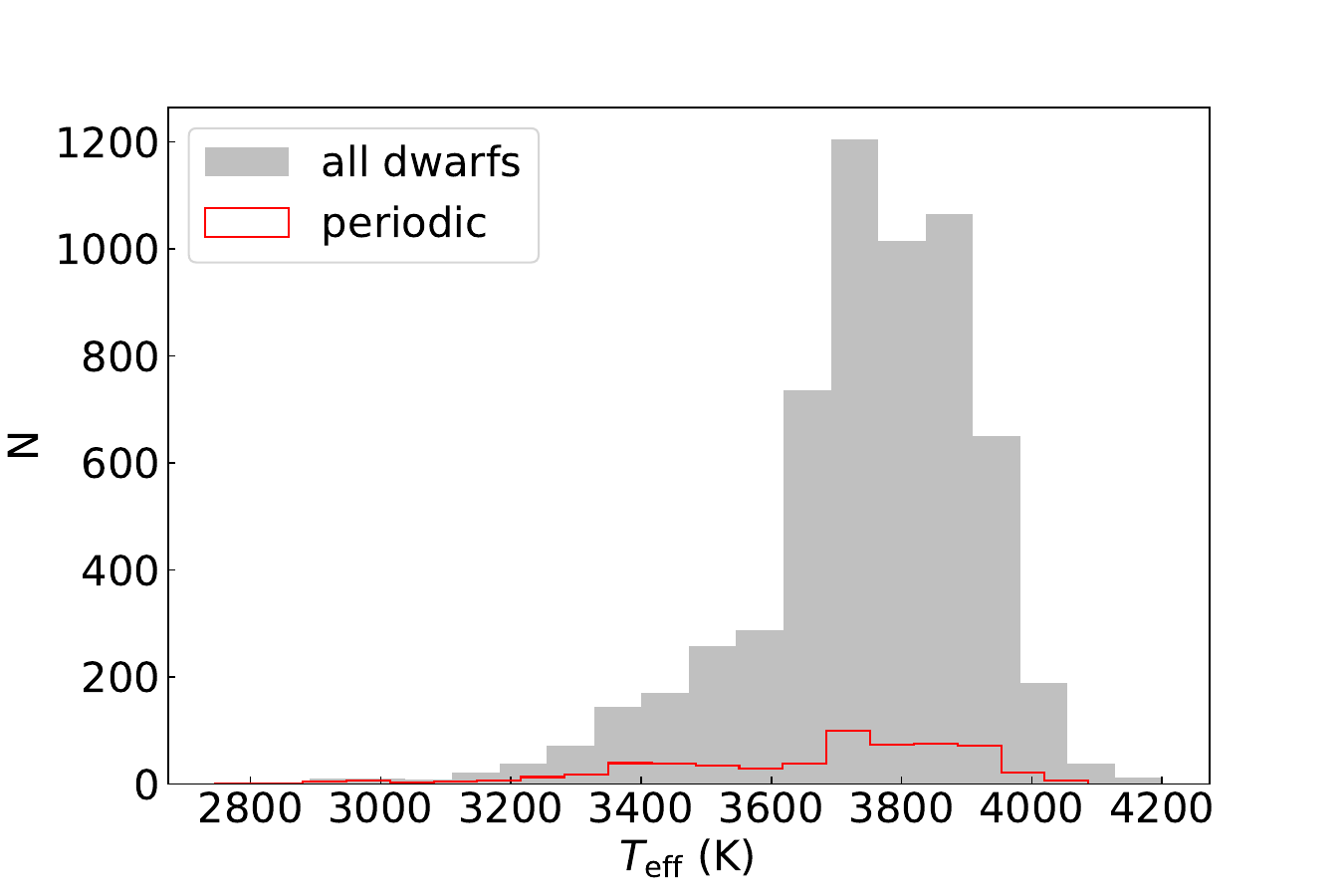}}
    \subfigure[]{
    \label{logl_all_p.fig}
    \includegraphics[width=0.49\textwidth]{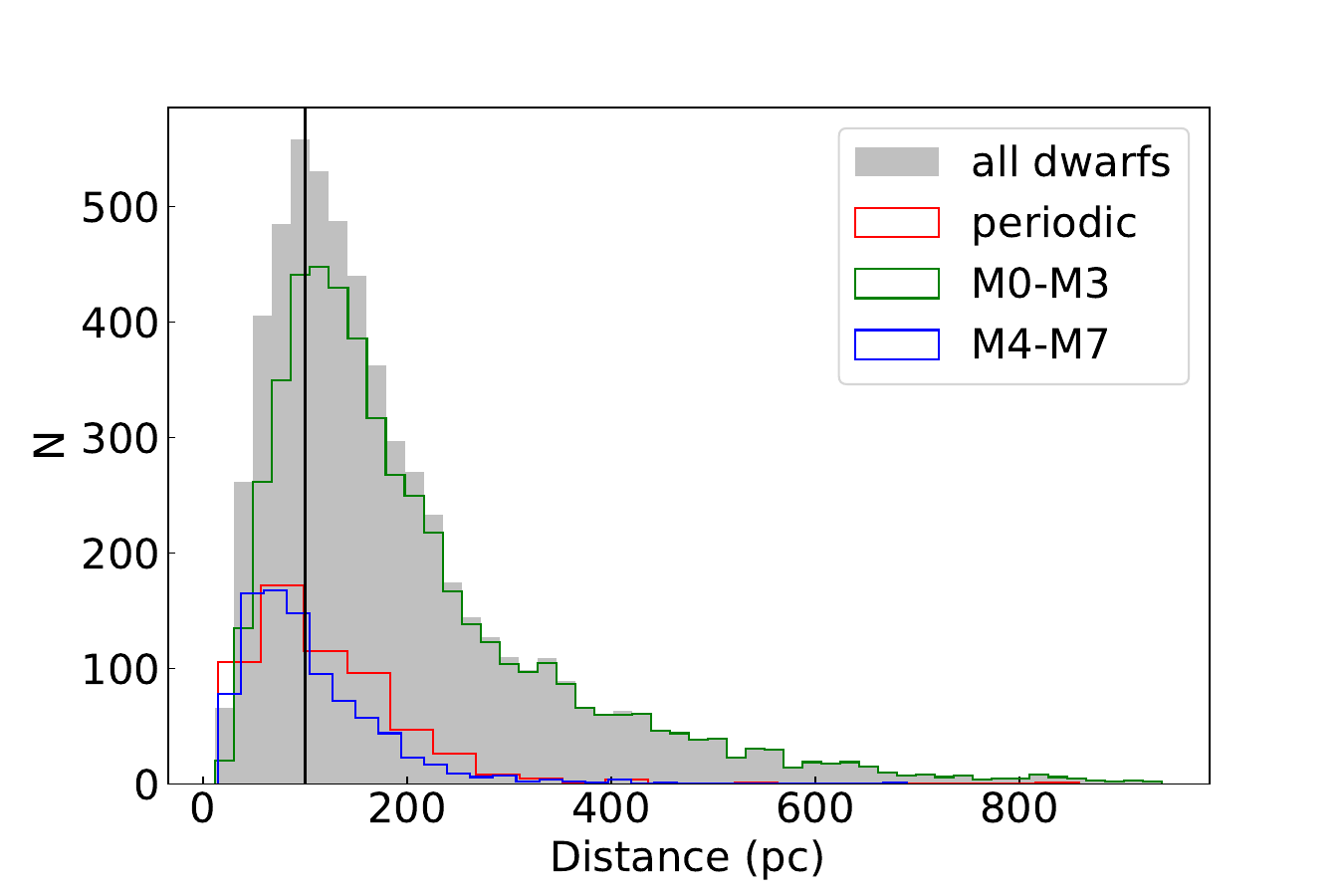}}
    \caption{Histograms of temperature (Panel a) and distance (Panel b) for the all dwarfs  and the periodic stars. Panel b also plots the distance distributions of M0--M3 and M4--M7 dwarfs. The vertical line marks a distance of 100 pc.}
    \label{distribution_par.fig}
\end{figure*}

Figure \ref{distribution_par.fig} shows the histograms of the effective temperature and distance for the sample stars, and Table \ref{all_pars.tab} shows the parameters of the stars.
The periodic sample refers to stars with rotational period estimations (Section \ref{rossby.sec}).
\begin{table*}[]
    \begin{center}
    \setlength{\tabcolsep}{8pt}
    \caption{Stellar parameters of the sample sources. $m_{\rm NUV}$ and $m_{\rm FUV}$ are the observed magnitude for GALEX NUV and FUV bands.}
    \begin{tabular}{cccccccccc}
    \hline
    Gaia id & RA. & DEC. & $T_{\rm eff}$ & log$g$ & [Fe/H] & Distance & E(B-V) & $m_{\rm NUV}$ & $m_{\rm FUV}$ \\
    \hline
    3204245546431688192 & 64.22870 & -3.16353 & 4024.66 & 5.07 & -0.28 & 251.17 & 0.05 & 18.31 & 18.36 \\
4437528912804247680 & 242.10202 & 4.60833 & 3901.46 & 5.09 & -0.21 & 200.81 & 0.05 & 22.52 & 23.35 \\
2485413805153218048 & 20.97025 & -1.42260 & 3864.15 & 5.12 & -0.18 & 531.54 & 0.05 & 21.44 & 22.88 \\
919952552803891072 & 116.28023 & 38.00565 & 3830.86 & 5.06 & -0.21 & 390.81 & 0.05 & 23.77 & 24.16 \\
2807530137536007680 & 8.06760 & 25.40596 & 3925.00 & 4.97 & -0.38 & 245.17 & 0.03 & 19.16 & 19.61 \\
136899784754020224 & 44.58209 & 33.72791 & 3914.89 & 5.12 & -0.36 & 313.34 & 0.10 & 19.89 & 20.94 \\
931687674766627200 & 125.17405 & 48.30424 & 3864.97 & 5.02 & -0.31 & 275.03 & 0.06 & 19.32 & 19.62 \\
3702591288979144192 & 192.98948 & 2.01242 & 3888.94 & 4.90 & -0.58 & 240.49 & 0.04 & 22.58 & 22.93 \\
2645837194505938432 & 351.61309 & 2.03311 & 3671.00 & 4.61 & -0.60 & 264.04 & 0.01 & 20.09 & 20.40 \\
888816067133565056 & 102.74665 & 30.70783 & 3880.45 & 5.10 & -0.20 & 361.34 & 0.06 & 21.82 & 22.55 \\
663532010117882240 & 123.94805 & 19.55912 & 3848.31 & 4.90 & -0.31 & 366.63 & 0.05 & 22.19 & 23.57 \\
3267315693766344704 & 47.25408 & 1.82766 & 4015.03 & 5.50 & 0.25 & 565.48 & 0.11 & 20.66 & 21.70 \\
3842831106588491008 & 136.54439 & -0.16756 & 3678.77 & 4.89 & -0.41 & 227.27 & 0.02 & 23.00 & 23.52 \\
800590429486863104 & 144.42856 & 37.71175 & 3904.64 & 5.02 & -0.28 & 312.39 & 0.05 & 20.77 & 21.96 \\
342617001561387904 & 27.75182 & 37.22069 & 3755.53 & 4.95 & -0.35 & 191.36 & 0.03 & 21.90 & 21.81 \\
17728461062321536 & 51.40735 & 13.77185 & 3896.48 & 4.98 & -0.32 & 301.21 & 0.27 & 19.02 & 18.53 \\
1591488350438997632 & 221.87057 & 48.69225 & 4030.27 & 5.09 & -0.33 & 288.57 & 0.04 & 18.89 & 19.09 \\
661934557160165376 & 132.71120 & 21.36956 & 3860.80 & 4.97 & -0.36 & 322.04 & 0.04 & 23.64 & 23.81 \\
3646602125373348864 & 213.76290 & -2.38854 & 3755.62 & 4.68 & -0.79 & 303.94 & 0.07 & 22.70 & 23.38 \\
3647017951221678976 & 212.87110 & -2.30502 & 4018.31 & 5.17 & -0.17 & 189.02 & 0.07 & 23.36 & 23.35 \\
    \hline
    \end{tabular}
    \label{all_pars.tab}
        \end{center}
    {NOTE. (This table is available in its entirety in machine-readable and Virtual Observatory (VO) forms in the online journal. A portion is shown here for guidance regarding its form and content.)}
\end{table*}

\section{Magnetic activity of M-type stars}
\label{activity.sec}

\subsection{UV activity indices}
\label{index.sec}
The definition of UV activity index is as following \citep{2011AJ....142...23F,2013MNRAS.431.2063S,2018ApJS..235...16B},
\begin{equation}
\label{activity_index.eq}
    R^{\prime}_{\rm UV} = \frac{f_{\rm UV,exc}}{f_{\rm bol}} = \frac{f_{\rm UV,obs}\times(\frac{d}{R})^2-f_{\rm UV,ph}}{f_{\rm bol}}
\end{equation}
where ``UV" stands for the NUV band or FUV band. 
The superscript ($^\prime$) means that the the UV emission from photosphere has been subtracted.
The $f_{\rm UV,exc}$ is the UV excess flux attributed to magnetic activity; $f_{\rm UV,obs}$ is the extinction-corrected flux inferred from the observed FUV or NUV magnitudes and extinction values; $f_{\rm UV,ph}$ is the photospheric flux from the stellar surface derived from stellar models;  $f_{\rm bol}$ is the bolometric flux calculated from effective temperatures following $f_{\rm bol} = \sigma_{B} T^{4}_{\rm eff}$.

We estimated the extinction-corrected UV flux following\footnote {\footnotesize https://asd.gsfc.nasa.gov/archive/galex/FAQ/counts\_\\background.html}
\begin{equation}
\begin{split}
\label{fuv_obs.eq}
    f_{\rm FUV,obs} = 10^{-0.4\times(m_{\rm FUV}-18.82-R_{\rm FUV} \times E(B-V))} \\
    \times 1.4 \times 10^{-15} \times \delta \lambda_{\rm FUV}
    \end{split}
\end{equation}
 and
 \begin{equation}
 \begin{split}
 \label{nuv_obs.eq}
    f_{\rm NUV,obs} = 10^{-0.4\times(m_{\rm NUV}-20.08-R_{\rm NUV} \times E(B-V))} \\
    \times 2.06 \times 10^{-16} \times \delta \lambda_{\rm NUV},
    \end{split}
\end{equation}
where $m_{\rm \lambda}$ is the observed magnitude from GALEX catalog, and $\delta \lambda_{\rm FUV}$ and $\delta \lambda_{\rm NUV}$ are the effective bandwidths of the FUV (442 \AA) and NUV (1060 \AA) filters, respectively \citep{2007ApJS..173..682M,2011AJ....142...23F}.
The extinction coefficients for FUV and NUV bands were calculated as 8.11 and 8.71 according to \cite{1989ApJ...345..245C}. 
The reddening $E(B-V)$ was derived from the Pan-STARRS DR1 (PS1) 3D dust map \citep{2015ApJ...810...25G} with $E(B-V) =0.884 \times {\rm Bayestar19}$.
For sources without extinction estimation from the PS1 dust map, we used the SFD dust map \citep{1998ApJ...500..525S} with $E(B-V) =0.884 \times E(B-V)_{\rm SFD}$ as a complement and only kept the sources with $E(B-V)$ $<$ 0.1.

The photospheric flux density from stellar surface was derived using BT-Settl (AGSS2009) stellar spectra models\footnote{http://svo2.cab.inta-csic.es/theory/newov2/}. 
The models include a 11-point grid of metallicities, with [Fe/H]= -4, $-$3.5, $-$3, $-$2.5, $-$2, $-$1.5, $-$1, $-$0.5, 0, 0.3, 0.5.
For each star, we first selected the models with two closest metallicities, and then extracted the best model by comparing the observed and theoretical log$T_{\rm eff}$ and log$g$ values for each metallicity. 
With the flux densities given by each model, we obtained the final flux density by linear interpolation using metallicity.
The photospheric flux was calculated by multiplying the flux density (in unit of erg/cm$^{2}$/s/\AA) with $\delta \lambda_{\rm UV}$ mentioned above.

The stellar radius was calculated from observed 2MASS ($J$, $H$, and $K_{\rm S}$) magnitudes, distance, extinction, and bolometric correction (BC):
\begin{equation}
\label{radius.eq}
\begin{split}
R&=\sqrt{\frac{L_{\rm bol}}{4\pi\sigma_{B}T^{4}}}\\
&=\sqrt{\frac{10^{(-0.4 \times (M_{\rm bol}-M_{\odot}))}L_{\odot}}{4\pi\sigma_{B}T^{4}}}\\
&=\sqrt{\frac{10^{(-0.4 \times (m_{\rm \lambda}-5{\rm log}d + 5 - A_{\rm \lambda} + BC-M_{\odot}))}L_{\odot}}{4\pi\sigma_{B}T^{4}}},
\end{split}
\end{equation}
where $M_{\odot}$ is solar bolometric magnitude (4.74 mag) and $L_{\odot}$ is solar bolometric luminosity (3.828$\times10^{33}$ erg/s).
$m_{\rm \lambda}$ is the apparent magnitude of $J$ band, $H$ band, or $K_{\rm S}$ band. 
The extinction $A_{\rm \lambda}$ is calculated as $A_{\lambda} = R_{\lambda} \times E(B-V)$, with the extinction coefficients estimated from \cite{1989ApJ...345..245C}.
The BC was derived using the {\it isochrones} Python module \citep{2015ascl.soft03010M}, with the stellar temperature, surface gravity, metallicity as inputs.
The final radius is obtained as the average value of the radii derived from different bands.
The results of the activity indices calculation are shown in Table \ref{all_results.tab}.

\begin{table*}[]
     \begin{center}
\caption{Stellar activity indices of the NUV, FUV, Ca \scriptsize{\uppercase\expandafter{\romannumeral2}} \normalsize H$\&$K and $\rm H\alpha$ bands of the total sample.}
\label{all_results.tab}
    \begin{tabular}{ccccccccc}
    \hline
    Gaia id & log$R^{\prime}_{\rm NUV}$ & log$R^{\prime}_{\rm FUV}$ & log$R^{\prime}_{\rm HK}$  & log$R^{\prime}_{\rm H\alpha}$ \\
    \hline
2772804845911842944 & $ -4.12 \pm 0.04 $ & $ -4.93 \pm 0.12 $ & $ -4.31 \pm 0.01 $ & $ -4.71 \pm 0.07 $ \\
710834787049323392 & $ -3.37 \pm 0.15 $ & $ -3.92 \pm 0.19 $ & $ -3.83 \pm 0.10 $ & $ -3.72 \pm 0.12 $ \\
680846672554039936 & $ -3.39 \pm 0.06 $ & $ -3.85 \pm 0.14 $ & $ -3.79 \pm 0.20 $ & $ -3.60 \pm 0.29 $ \\
3146147042781951744 & $ -3.46 \pm 0.08 $ & $ -4.05 \pm 0.16 $ & $ -4.02 \pm 0.00 $ & $ -3.80 \pm 0.05 $ \\
4413290007171520768 & $ -4.13 \pm 0.05 $ & $ -4.85 \pm 0.18 $ & $ -4.50 \pm 0.00 $ & $ -4.63 \pm 0.13 $ \\
3367511306484702592 & $ -3.41 \pm 0.11 $ & $ -3.78 \pm 0.16 $ & $ -4.02 \pm 0.02 $ & $ -3.90 \pm 0.07 $ \\
136899784754020224 & $ -2.00 \pm 0.04 $ & $ -2.50 \pm 0.08 $ & $ -4.05 \pm 0.09 $ & $ -3.94 \pm 0.28 $ \\
1038927991625880832 & $ -4.06 \pm 0.02 $ & $ -5.22 \pm 0.14 $ & $ -4.42 \pm 0.00 $ & $ -5.45 \pm 0.52 $ \\
578274091791263104 & $ -3.37 \pm 0.02 $ & $ -4.14 \pm 0.09 $ & $ -4.41 \pm 0.02 $ & $ -3.87 \pm 0.04 $ \\
1303190835359164416 & $ -3.57 \pm 0.04 $ & $ -4.57 \pm 0.15 $ & $ -3.97 \pm 0.01 $ & $ -4.08 \pm 0.06 $ \\
3809500549759137152 & $ -3.70 \pm 0.04 $ & $ -4.49 \pm 0.13 $ & $ -4.18 \pm 0.02 $ & $ -3.93 \pm 0.03 $ \\
312984407278269568 & $ -3.31 \pm 0.15 $ & $ -3.46 \pm 0.21 $ & $ -4.31 \pm 0.01 $ & $ -4.46 \pm 0.09 $ \\
663532010117882240 & $ -2.73 \pm 0.12 $ & $ -3.34 \pm 0.17 $ & $ -4.89 \pm 0.01 $ & $ -4.99 \pm 0.79 $ \\
2593114821680113536 & $ -3.54 \pm 0.04 $ & $ -4.16 \pm 0.08 $ & $ -3.93 \pm 0.06 $ & $ -3.75 \pm 0.10 $ \\
673218020362903040 & $ -3.57 \pm 0.17 $ & $ -3.65 \pm 0.21 $ & $ -4.10 \pm 0.03 $ & $ -3.91 \pm 0.16 $ \\
3844876610533210496 & $ -3.68 \pm 0.05 $ & $ -4.46 \pm 0.10 $ & $ -4.34 \pm 0.02 $ & $ -4.30 \pm 0.04 $ \\
100307419305322496 & $ -3.27 \pm 0.08 $ & $ -3.81 \pm 0.14 $ & $ -3.68 \pm 0.05 $ & $ -3.38 \pm 0.02 $ \\
685749635420118912 & $ -3.80 \pm 0.10 $ & $ -4.26 \pm 0.18 $ & $ -4.87 \pm 0.01 $ & $ -4.95 \pm 0.87 $ \\
\hline
    \end{tabular}
     \end{center}
        {NOTE. (This table is available in its entirety in machine-readable and Virtual Observatory (VO) forms in the online journal. A portion is shown here for guidance regarding its form and content.)}
\end{table*}

The $\delta \lambda_{\rm FUV}$ and $\delta \lambda_{\rm NUV}$ in Eq. \ref{fuv_obs.eq} and \ref{nuv_obs.eq} are 442 \AA\ and 1060 \AA, respectively \citep{2007ApJS..173..682M, 2011AJ....142...23F}.
We noticed that some studies estimated UV flux using narrower bandwidths, specifically $\delta \lambda_{\rm FUV} = 268$ \AA\ and $\delta \lambda_{\rm NUV} = 732$ \AA\ \citep{2013MNRAS.431.2063S,2018ApJS..235...16B}. 
In the latter case, both the observed flux and the photospheric flux decrease, leading to smaller activity indices (i.e., a reduction of $\approx$0.2 in $R^{\prime}_{\rm FUV}$ and $\approx$0.16 in $R^{\prime}_{\rm NUV}$).

\subsection{Chromospheric activity indices}
\label{chromospheric.sec}

\subsubsection{\rm{Ca \scriptsize{\uppercase\expandafter{\romannumeral2}} \normalsize H$\&$K} \emph{activity index}}
\label{hk_activity.sec}

We calculated the S-index and then obtained the Ca \scriptsize{\uppercase\expandafter{\romannumeral2}} \normalsize H$\&$K activity index $R^{\prime}_{\rm{HK}}$ using the low-resolution spectra from LAMOST DR9.

We calculated the S-index and $R'_{\rm HK}$ based on Yang et al. 2023 (in prep.). Here we give a brief description of the method. The LAMOST S-index ($S_{\rm LAMOST}$) can be written as \citep{2011arXiv1107.5325L,2016NatCo...711058K,2017A&A...600A..13A}
\begin{equation}
\begin{aligned}
 S_{\rm LAMOST}&=8\alpha \cdot \frac{\Delta \lambda_{\rm H}}{\Delta \lambda_{\rm R}} \cdot \frac{\tilde{f}_{\rm H} +\tilde{f}_{\rm K}}{\tilde{f}_{\rm V} +\tilde{f}_{\rm R}}\\ &= 8\alpha \cdot \frac{1.09 \rm \AA}{20 \rm \AA} \cdot \frac{\tilde{f}_{\rm H} +\tilde{f}_{\rm K}}{\tilde{f}_{\rm V} +\tilde{f}_{\rm R}}
 \end{aligned}
 \end{equation}
where $\tilde{f}_{\rm H}, \tilde{f}_{\rm K}, \tilde{f}_{\rm V}, \tilde{f}_{\rm R}$ are the  mean flux per wavelength interval in four bandpasses, and the correction factor $\alpha$ is 2.4 \citep{1991ApJS...76..383D}.

A linear calibrating equation between $S_{\rm LAMOST}$ and the Mount Wilson Observatory scale ($S_{\rm MWO}$) was derived through the common stars of the LAMOST spectra and \cite{2018A&A...616A.108B}, which was found (Yang et al. 2023, in prep.) to be
\begin{equation}\label{eq_sc}
\begin{aligned}
 S_{\rm MWO}&=a \cdot S_{\rm LAMOST} + b\\
 a&= 2.89^{+0.21}_{-0.17}\\
 b&= -0.65^{+0.18}_{-0.20}
 \end{aligned}
 \end{equation}

We then converted the calibrated S-index to $R'_{\rm HK}$ following the relation in terms of the color index $B-V$ \citep[see][for more details]{2017A&A...600A..13A}, while the $B-V$ index was derived from effective temperature and metallicity based on the Dartmouth Stellar Evolutionary Database \citep{2008ApJS..178...89D}.

\subsubsection{$\rm H\alpha$ activity index}
\label{ha_activity.sec}

We calculated the stellar activity index $R^{\prime}_{\rm{H\alpha}}$ with $\rm{H\alpha}$ equivalent width (EW) using the LAMOST DR9 low-resolution spectra. 
$R^{\prime}_{\rm{H\alpha}}$ is the normalized $\rm{H\alpha}$ luminosity defined as \citep{2004PASP..116.1105W}
\begin{equation}
\label{Rha}
R_{\rm{H\alpha}}^{'} = \chi \times \rm{EW}^{'}.
\end{equation}
Here $\rm{EW}^{\prime}$ is the EW caused by dynamo-driven magnetic activity, which was calculated as follows,
\begin{equation}
\label{ew+}
\rm{EW}^{\prime} = \rm{EW} - \rm{EW}_{\rm{basal}},
\end{equation}
where
\begin{equation}
\label{EW}
\rm{EW}_{\rm{H\alpha}} = \int \frac{\emph{F}_{\rm{c}} - \emph{F}_{\rm{\lambda}}}{\emph{F}_{\rm{\lambda}}}.
\end{equation}
$\rm{EW}_{\rm{basal}}$ represents the $\rm{H\alpha}$ emission due to physical processes unrelated to magnetic activity. The basal flux of $\rm{H\alpha}$ was derived through fitting a spline function to the $EWs$ of the most inactive stars in our sample.

The normalized factor $\chi$ was estimated following \citep{2023ApJS..264...12H}
\begin{equation}
\label{chi}
\chi = \frac{f_{\rm{\lambda}6564}}{f_{\rm{bol}}} = \frac{f_{\rm{\lambda}6564}}{\sigma_{B} T^{4}},
\end{equation}
where $f_{\rm{\lambda}6564}$ was the continuum flux at 6564 \AA, which was derived through fitting the continuum of PHOENIX model spectrum \citep{2013A&A...553A...6H} corresponding to the stellar parameters of our targets, and $f_{\rm{bol}}$ was the bolometric flux on stellar surface \citep[for  more details refer to][]{2023ApJS..264...12H}.

\subsection{Rotation periods and Rossby number}
\label{rossby.sec}

\begin{figure}[htp]
   \center
   \includegraphics[width=0.45\textwidth]{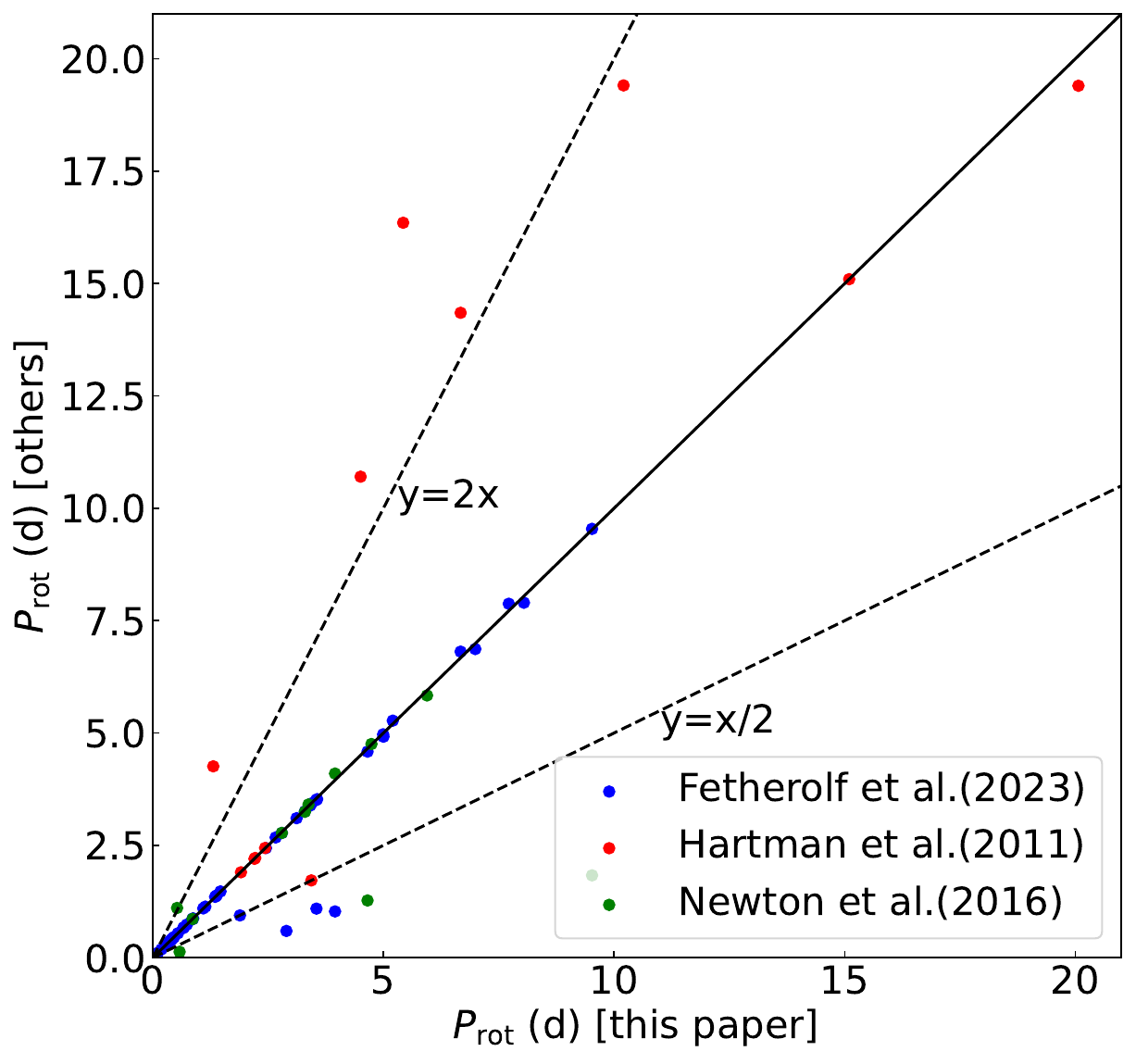}
   \caption{Comparison between the periods from TESS light curve data (in this paper) and the periods derived from previous studies. The blue points are the periods from \cite{2023ApJS..268....4F}, the red points are the periods from \cite{2011AJ....141..166H}, and the green points are the periods from \cite{2016ApJ...821...93N}. The black line is y=x, and the black dashed lines are y=2x and y=x/2, respectively.}
   \label{p_compare.fig}
\end{figure}

We first collected stellar rotation periods from previous photometric surveys, including \emph{Kepler/K2}, ASAS-SN, ZTF, etc. 
The rotation periods of 199, 48, 76, 259 and 1 stars were obtained from the \emph{K2}, ZTF, ASAS-SN, TESS and HATNet data, respectively.
For these sources, we downloaded the photometric data and and visually checked the phase-folded light curves.
Next, we cross-matched our sample with the TESS archive data, downloaded the light curve data, and estimated the rotation periods using the Lomb-Scargle method \citep{2018ApJS..236...16V}.
The phase-folded curves were checked by eye as well.
Our period estimations are in agreement with those from \cite{2016ApJ...821...93N} and \cite{2023ApJS..268....4F}, indicating the reliability of our period determination, although some periods given by \cite{2011AJ....141..166H} are doubled compared with our estimations (Figure \ref{p_compare.fig}).

The Rossby number is usually used to trace the stellar rotation, which is defined as the ratio of the rotation period to the convective turnover time (Ro $=$ P/$\tau$).
We obtained the $\tau$ value using a grid of stellar evolution models from the Yale-Potsdam Stellar Isochrones \citep[YaPSI;][]{2017ApJ...838..161S} following \cite{2020ApJ...902..114W}.
Here we used the effective temperature $T_{\rm eff}$ and bolometric luminosity $L_{\rm bol}$ to fit the model evolutionary tracks.
For each star, we obtained best-fit models for close metallicities, and calculated the final $\tau$ value by linear interpolation to the metallicity.
The period, convective turnover time and Rossby number of stars are shown in Table \ref{all_p.tab}.
Note that compared with the classical empirical estimate of $\tau$ \citep{1984ApJ...287..769N}, the ratio between the $\tau$ from theoretical YAPSI model and empirical $\tau$ values  are around 3 \citep{2020ApJ...902..114W}.

\begin{table}[]
    \begin{center}
    \setlength{\tabcolsep}{2pt}
    \caption{The parameters of periodic sample.}
    \begin{tabular}{ccccc}
    \hline
    Gaia id & Period & Ref.\tnote{1} & $\tau$ & Ro \\
    \hline
688528547980217088 & 17.96000 & 1 & 88.71 & 0.20 \\
3132744335339029376 & 9.79997 & 4 & 82.72 & 0.12 \\
740411679900325120 & 6.10162 & 4 & 94.51 & 0.06 \\
740266991041907840 & 1.94940 & 3 & 96.13 & 0.02 \\
5181296851945039360 & 10.64136 & 3 & 111.88 & 0.10 \\
2481571802287972608 & 5.05442 & 4 & 86.33 & 0.06 \\
2700541845063604992 & 0.76202 & 3 & 92.00 & 0.01 \\
2651027782742343296 & 31.01000 & 1 & 105.75 & 0.29 \\
3866696762383440256 & 18.80000 & 1 & 96.85 & 0.19 \\
879903868258387968 & 17.96056 & 4 & 119.21 & 0.15 \\
679329621385921536 & 1.20710 & 3 & 98.81 & 0.01 \\
631716232416975488 & 17.07000 & 1 & 96.85 & 0.18 \\
3146147042781951744 & 2.08639 & 4 & 97.23 & 0.02 \\
638525198689503360 & 22.59000 & 1 & 109.31 & 0.21 \\
1415189357506518272 & 8.88880 & 2 & 99.60 & 0.09 \\
1325695295758670208 & 10.72802 & 4 & 107.98 & 0.10 \\
222248893824346112 & 10.05350 & 4 & 109.91 & 0.09 \\
127275106640433664 & 4.64074 & 4 & 104.41 & 0.04 \\
1904506975722625408 & 6.67416 & 4 & 90.47 & 0.07 \\
1896376289095348352 & 3.30535 & 4 & 102.00 & 0.03 \\
    \hline
    \end{tabular}
    \label{all_p.tab}
     \end{center}
     Note. The references include 
        1: \cite{2020A&A...635A..43R} from K2 data;
        2: \cite{2020ApJS..249...18C} from ZTF data;
        3: \cite{2018MNRAS.477.3145J} and \cite{2023MNRAS.519.5271C} from ASAS-SN data; 
        4: \cite{2022RNAAS...6...96H}, \cite{2022ApJS..258...16P} and our estimations from TESS data;
        5: \cite{2016ApJ...821...93N} from HATNet data. (This table is available in its entirety in machine-readable and Virtual Observatory (VO) forms in the online journal. A portion is shown here for guidance regarding its form and content.)
\end{table}

\subsection{Activity-rotation relationship}
\label{R_ro.sec}

\begin{figure*}
    \centering
    \subfigure[]{
    \label{nuv_ro.fig}
    \includegraphics[width=0.49\textwidth]{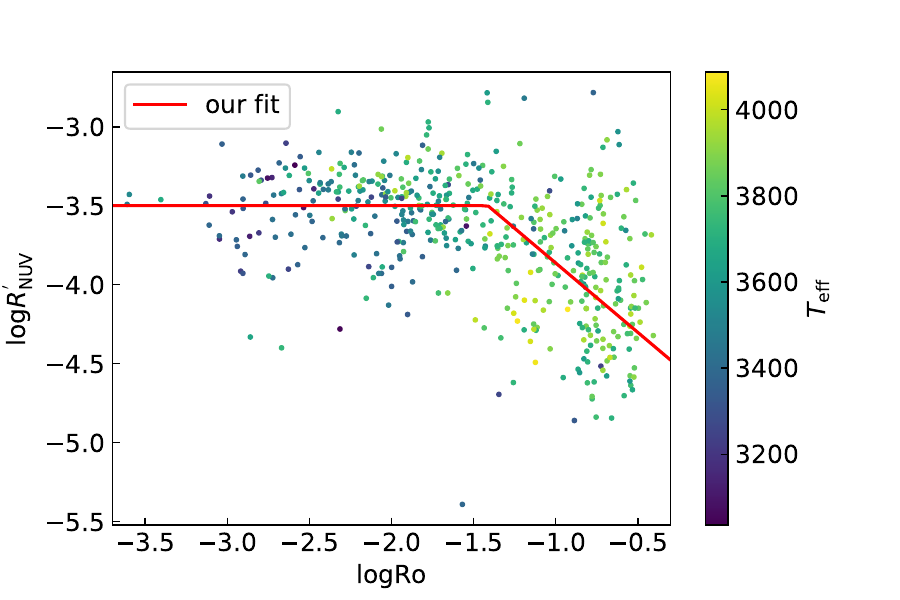}}
    \subfigure[]{
    \label{fuv_ro.fig}
    \includegraphics[width=0.49\textwidth]{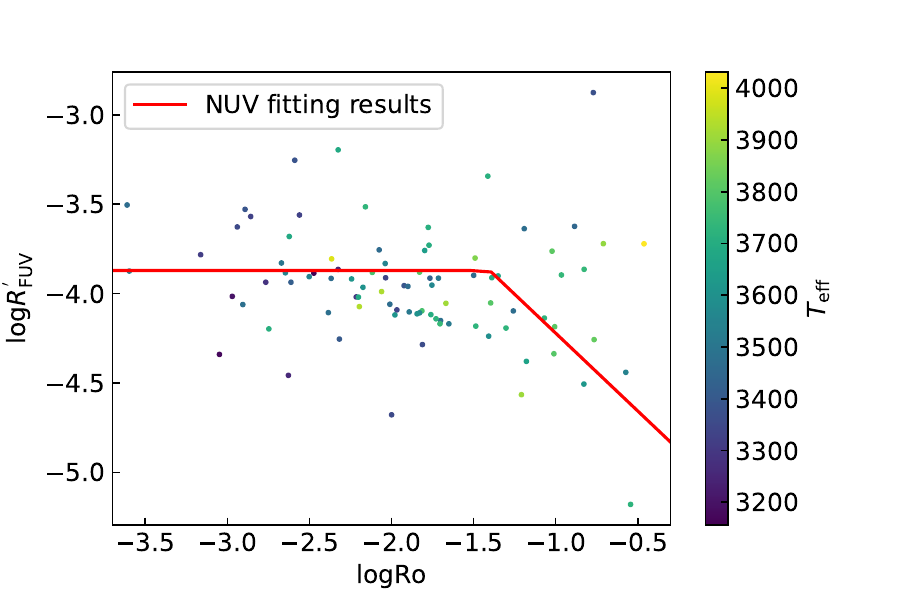}}
    \subfigure[]{
    \label{hk_ro.fig}
    \includegraphics[width=0.49\textwidth]{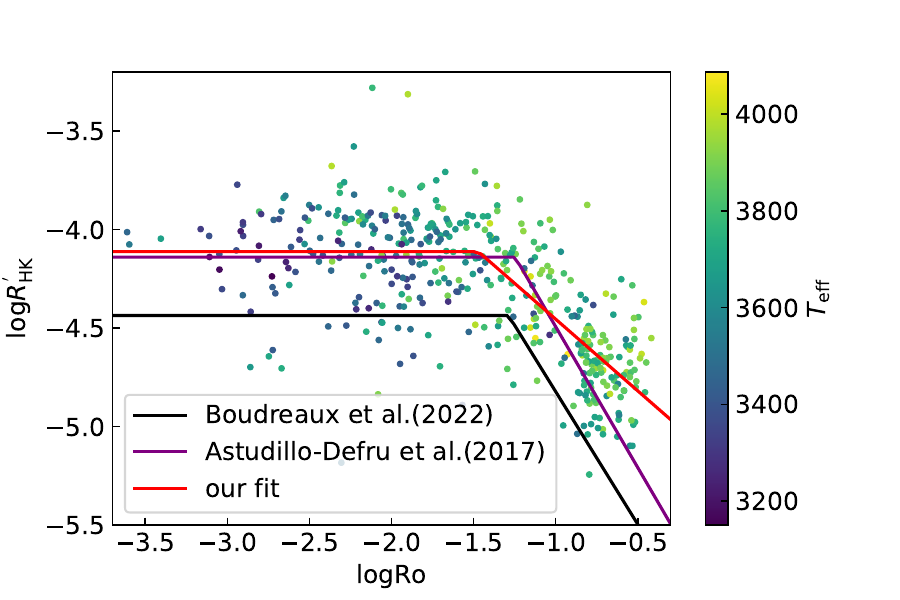}}
    \subfigure[]{
    \label{ha_ro.fig}
    \includegraphics[width=0.49\textwidth]{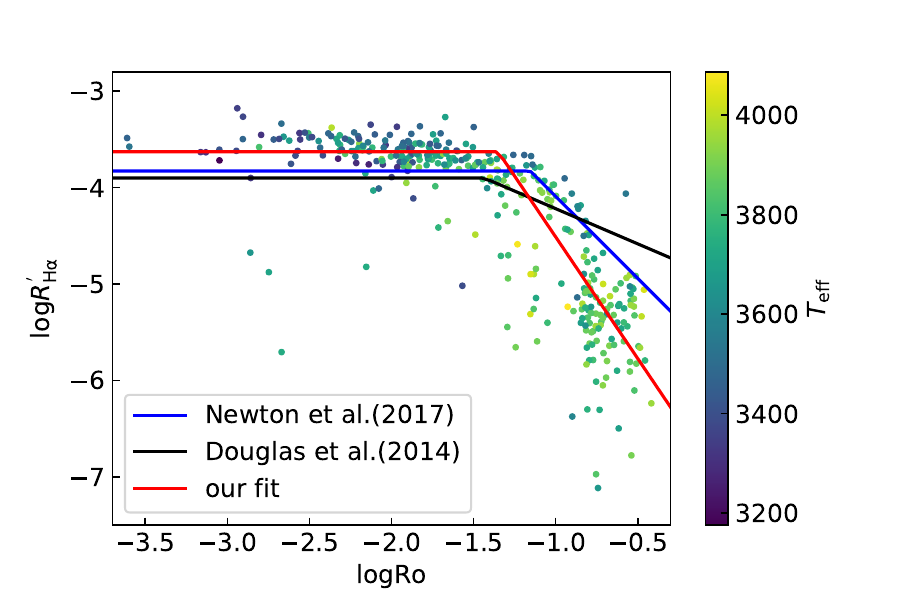}}
   \caption{The activity-rotation relation of the GAELX NUV band (Panel a), FUV band (Panel b), Ca \scriptsize{\uppercase\expandafter{\romannumeral2}} \normalsize H$\&$K line (Panel c), and H$\alpha$ line (Panel d), respectively. The points means the periodic sample, and the color bar represent the effective temperature. The black and purple line in Panel (c) correspond to the findings reported in \cite{2022ApJ...929...80B}, and \cite{2017A&A...600A..13A}, respectively, but they have been horizontally shifted to the left by a factor of $\rm Ro/3$. The blue and black line in Panel (d) are from \cite{2017ApJ...834...85N} and \cite{2014ApJ...795..161D}, respectively, but shifted to left with a ratio of $\rm Ro/3$. The red line represent the fitting result from our sample.}
   \label{r_ro.fig}
\end{figure*}

For the UV and chromospheric activity indices, we performed a piecewise fitting analysis to study the activity-rotation relationship following
\begin{equation}
\label{r_ro.eq}
 R^{\prime}_{\rm \lambda}=
\begin{cases}
 R^{\prime}_{\rm \lambda,sat},\quad \rm if\ Ro \le \rm Ro_{sat}\\
 C\times \rm Ro^{\beta},\quad \rm if\ Ro > \rm Ro_{sat}.
 \end{cases}
\end{equation}
Here $\lambda$ indicates the activity indices of NUV, FUV,  $\rm H\alpha$ and Ca \scriptsize{\uppercase\expandafter{\romannumeral2}} \normalsize H$\&$K lines.

\begin{figure}
   \center
   \includegraphics[width=0.48\textwidth]{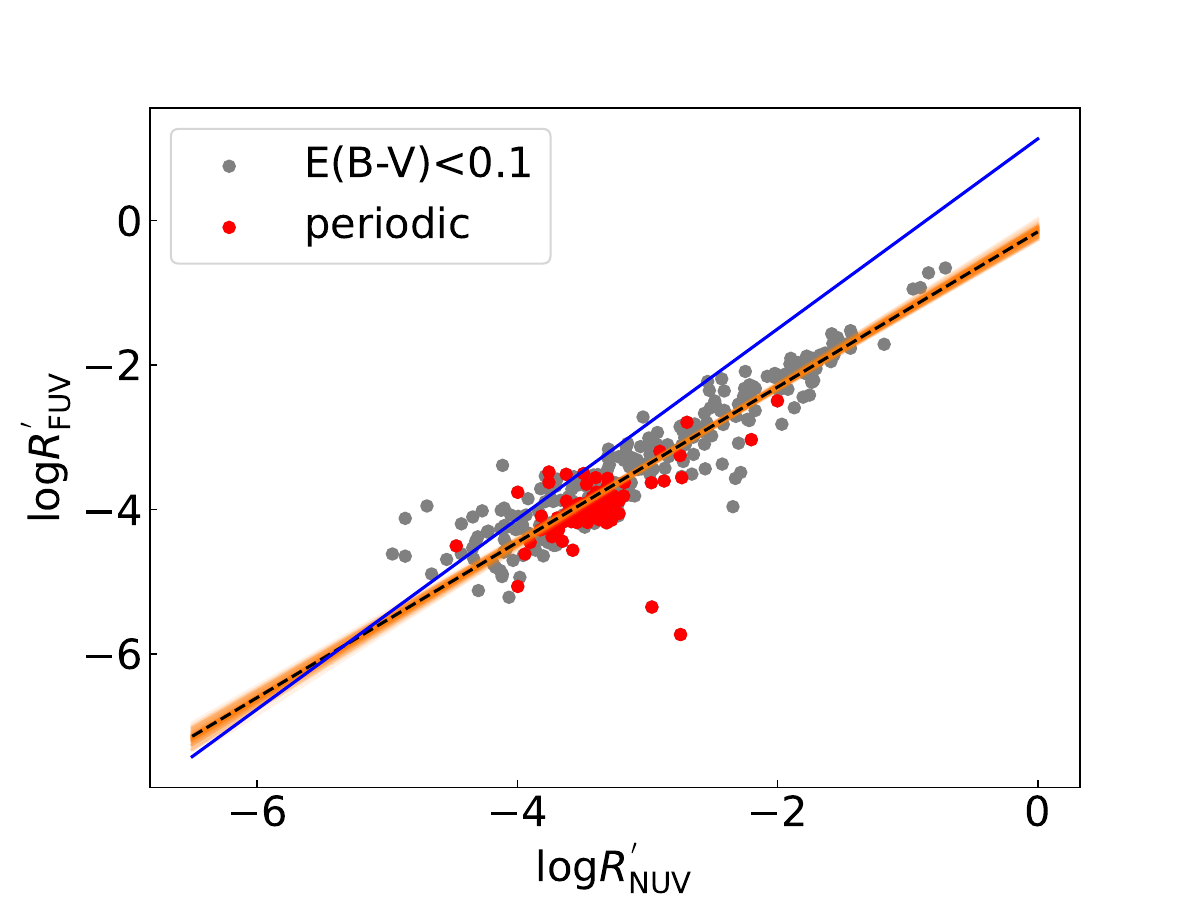}
   \caption{The compare of $R^{\prime}_{\rm NUV}$ and $R^{\prime}_{\rm FUV}$ in log-log scale. The gray points are the stars that E(B-V) less than 0.1, and the red points are the periodic sample. The black dashed line is the MCMC fitting result. The blue line is the results from \cite{2013MNRAS.431.2063S}.}
   \label{fuv_nuv.fig}
\end{figure}

Figure \ref{nuv_ro.fig} shows the fitting result for the NUV band, with
log$R^{\prime}_{\rm NUV,sat}=-3.50\pm 0.02$,  log$\rm Ro_{sat}=-1.40\pm 0.05$ and $\beta = -0.87\pm 0.07$ (Table \ref{ro_fit.tab}). 
Due to the limited number of samples with FUV observations, we did not directly perform a fitting analysis for the FUV band. 
However, we can derive the activity-rotation relation for FUV band by fitting a relation between $R^{\prime}_{\rm FUV}$ and $R^{\prime}_{\rm NUV}$. 
As shown in Figure \ref{fuv_nuv.fig}, there is a linear relationship between log$R^{\prime}_{\rm NUV}$ and log$R^{\prime}_{\rm FUV}$, which can be quantified as follows,
\begin{equation}
\label{fuv_nuv.eq}
    {\rm log} R^{\prime}_{\rm FUV} = (1.07^{+0.02}_{-0.02}) \times {\rm log}R^{\prime}_{\rm NUV} - (0.16^{+0.06}_{-0.06})
\end{equation}

The fitting result slightly differs from the result given by \cite{2013MNRAS.431.2063S}. We found the larger slope ($\approx$1.3) reported by \cite{2013MNRAS.431.2063S} is mainly caused by the M stars in the TW Hya association, which clearly shows a deviation in the $R_{\rm UV}$ with their 10 pc sample (see their Figure 14).
The relation between the FUV activity and rotation can then be described as log$R^{\prime}_{\rm FUV,sat}=-3.87\pm 0.02$, log$\rm Ro_{sat}=-1.40\pm 0.05$ and $\beta = -0.87\pm 0.07$.

\begin{table}[]
    \centering
    \caption{The fitting results of the activity-rotation relations for four proxies.}
    \label{ro_fit.tab}
    \begin{tabular}{cccc}
    \hline\noalign{\smallskip}
    band & log$R^{\prime}_{\rm \lambda,sat}$ & log$\rm Ro_{sat}$ & $\beta$ \\
    \hline\noalign{\smallskip}
    NUV & $-3.50\pm 0.02$ &  $-1.40\pm 0.05$ & $-0.87\pm 0.07$ \\
    FUV & $-3.87\pm 0.02$ &  $-1.40\pm 0.05$ & $-0.87\pm 0.07$ \\
    Ca \scriptsize{\uppercase\expandafter{\romannumeral2}} \normalsize H$\&$K & $-4.11\pm 0.02$ & $-1.47\pm 0.05$ & $-0.73\pm 0.06$ \\
    $\rm H\alpha$ & $-3.63\pm 0.03$ & $-1.35\pm0.03$ & $-2.52\pm 0.14$ \\
    \hline\noalign{\smallskip}
    \end{tabular}
\end{table}

Figure \ref{hk_ro.fig} shows the relation between log$R^{\prime}_{\rm{HK}}$ and $\rm Ro$. 
We fitted the relation using Equation \ref{r_ro.eq} and obtained the parameters as log$R^{\prime}_{\rm{HK, sat}}=-4.11\pm 0.02$, log$\rm Ro_{sat}=-1.47\pm 0/05$, and $\beta = -0.73\pm 0.06$. 
The $\beta$ value is in good agreement with that from \citet{2020NatAs...4..658L} ($\beta \approx -0.97$), who studied the Ca \scriptsize{\uppercase\expandafter{\romannumeral2}} \normalsize H$\&$K activity of F, G, and K type stars. 
The log$R^{\prime}_{\rm{HK, sat}}$ is notably higher than the value reported by \cite{2022ApJ...929...80B}, but very similar to \cite{2017A&A...600A..13A}. 
The log$\rm Ro_{sat}$ and $\beta$ values are different with those values from \citet{2017A&A...600A..13A} and \citet{2022ApJ...929...80B}. One explanation is that there are very few sources located in the transition region in their sample, and the choice of the knee point greatly affects the slope in the unsaturated region. 
In addition, although our sample has a larger number, there are also many scatters, which would affect the accuracy of the fitting results.

Figure \ref{ha_ro.fig} shows the best-fit result with log$R^{\prime}_{\rm H\alpha,sat}=-3.63\pm 0.03$, log$\rm Ro_{sat}=-1.35\pm 0.03$ and $\beta = -2.52\pm 0.14$. 
The $\beta$ value is different with previous studies, e.g., $\beta = -1.7\pm 0.1$ from \cite{2017ApJ...834...85N} or $\beta = -0.73^{+0.16}_{-0.12}$ from \cite{2014ApJ...795..161D}. In the unsaturated region, both of their samples have many sources with lower activity indices (log$R^{\prime}_{\rm H\alpha} \approx -6$) which do not match their relations. We tried to perform a fitting including these sources and derived a larger slope.
The slight difference of log$R^{\prime}_{\rm H\alpha,sat}$ is likely due to the different calculation method of the normalized factor $\chi$. Moreover, \cite{2014ApJ...795..161D} studied a mono-age population and didn't correct for baseline. The above‐mentioned factors may result in some differences in the activity distributions and fitted relations.

\subsection{Distribution of stellar activity indices}
\label{distribution.sec}

\begin{figure*}[htp]
    \centering
    \subfigure[]{
    \label{Rnuv_ms_g.fig}
    \includegraphics[width=0.49\textwidth]{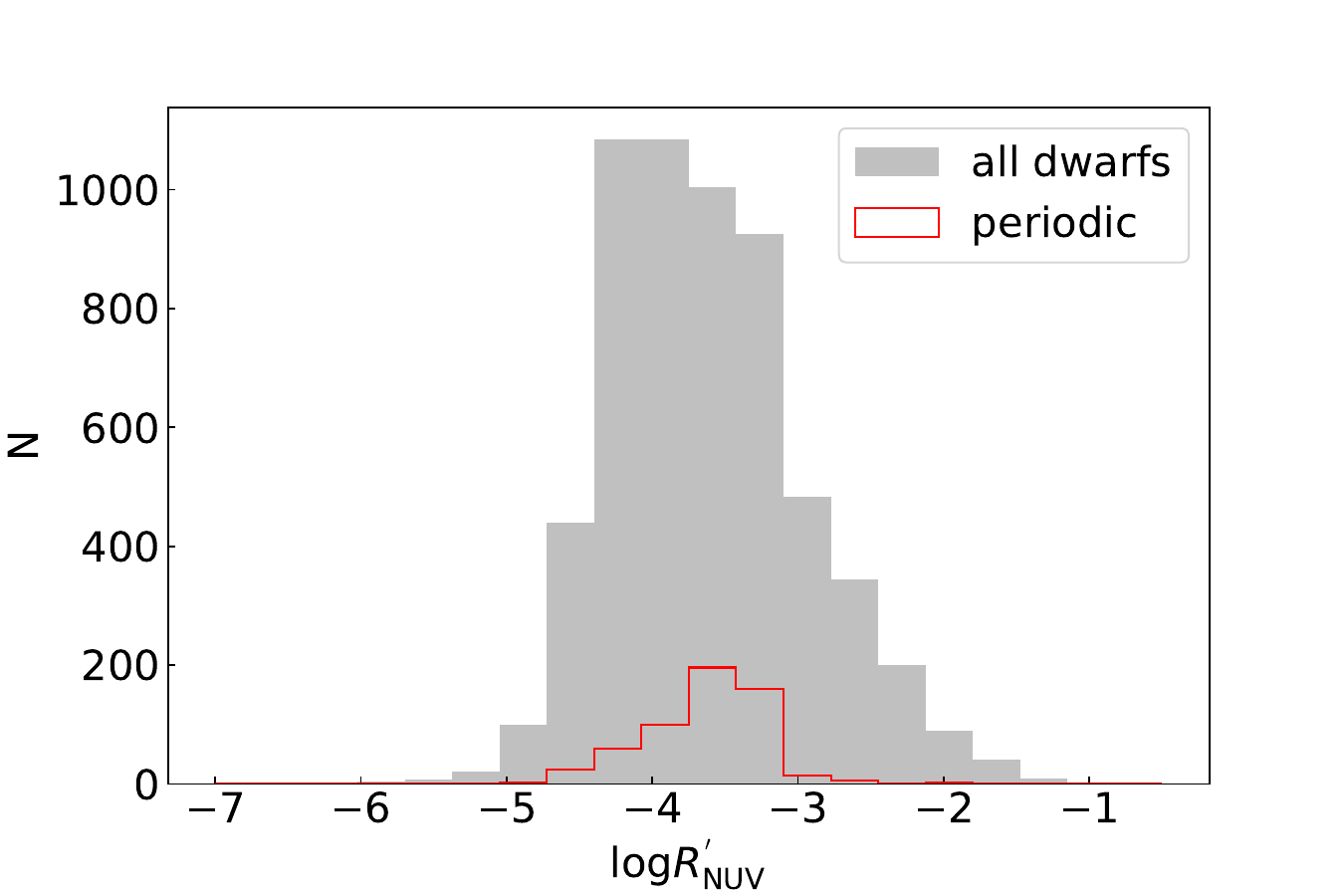}}
    \subfigure[]{
    \label{Rfuv_ms_g.fig}
    \includegraphics[width=0.49\textwidth]{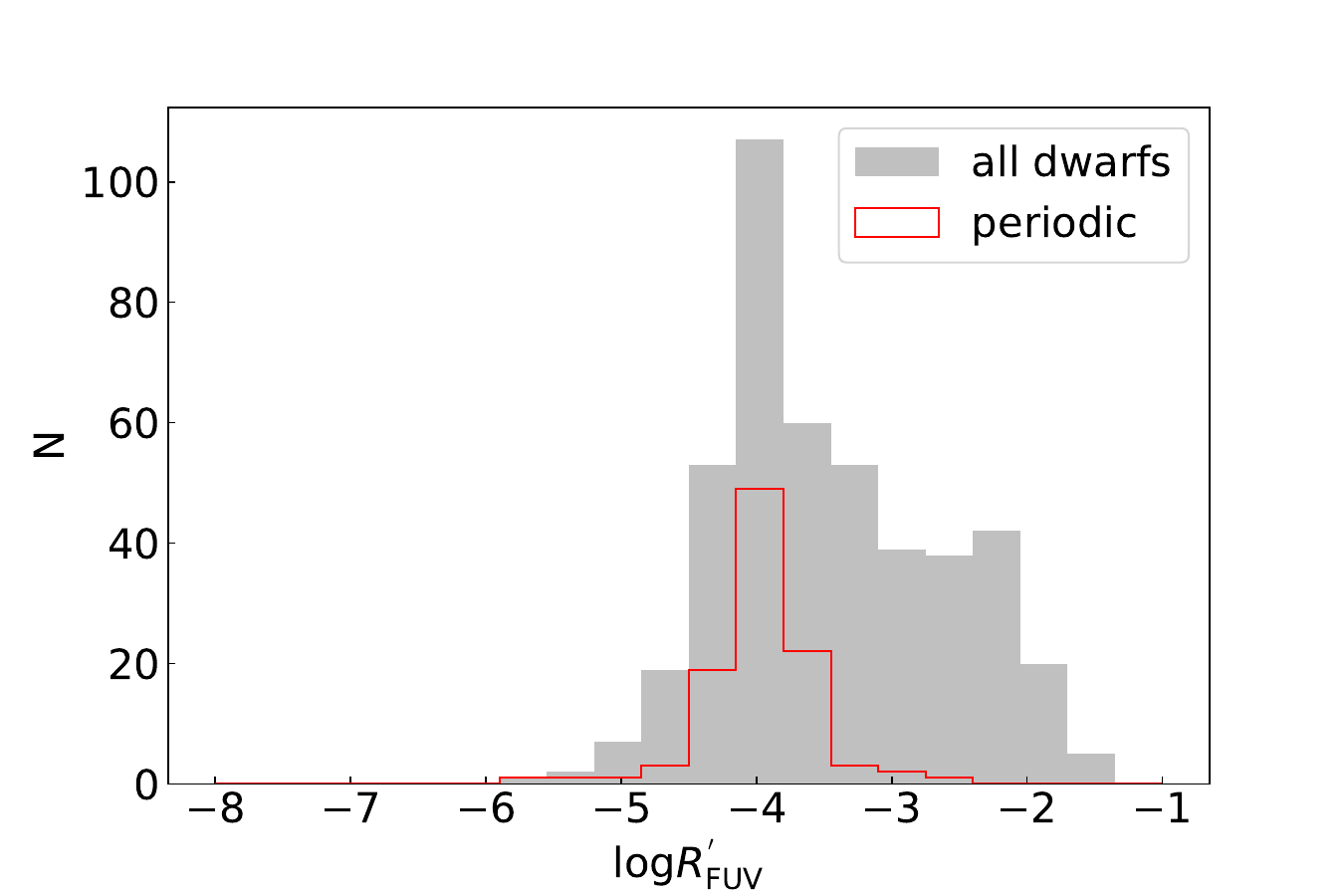}}
    \subfigure[]{
    \label{Rhk_ms_g.fig}
    \includegraphics[width=0.49\textwidth]{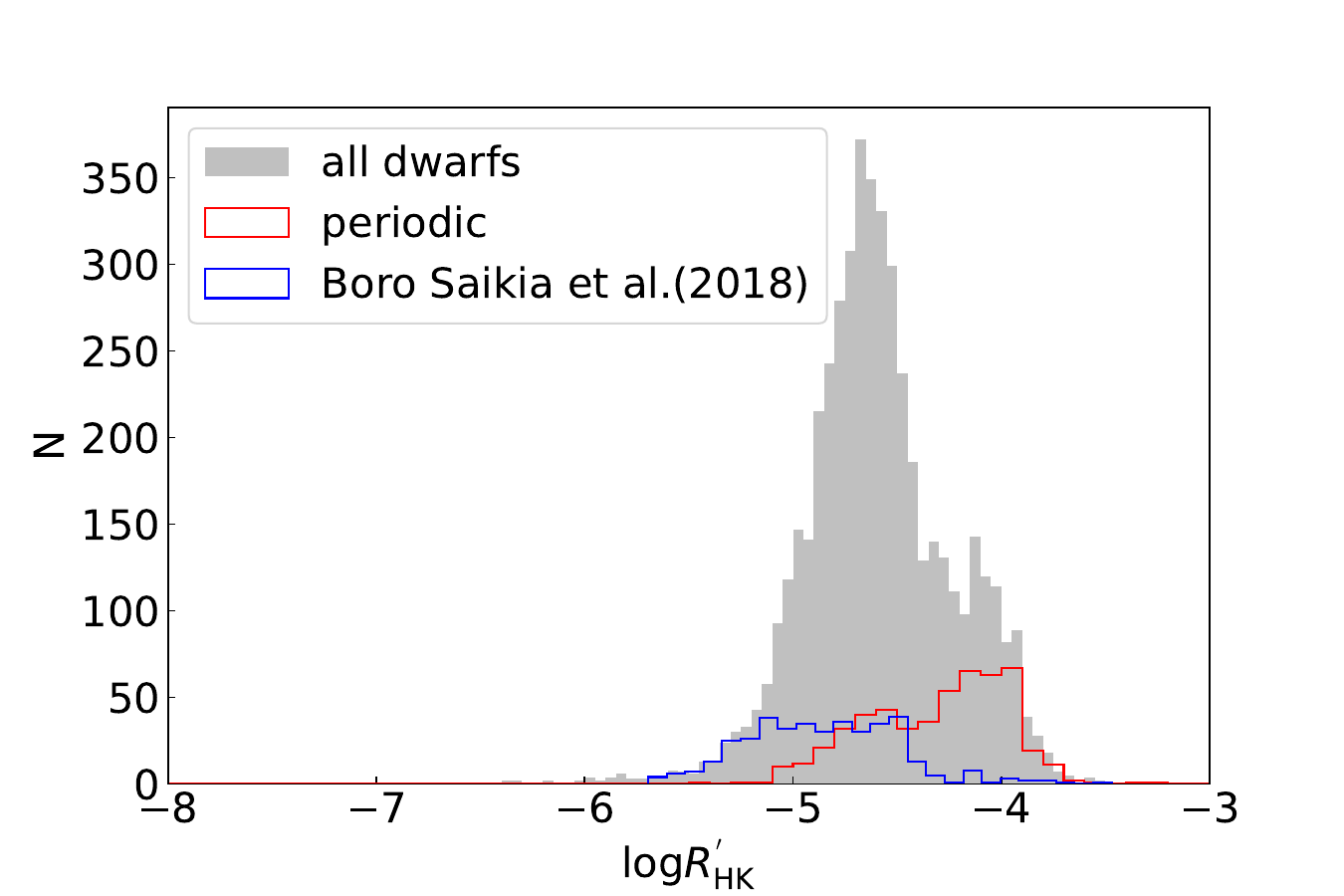}}
    \subfigure[]{
    \label{Rha_ms_g.fig}
    \includegraphics[width=0.49\textwidth]{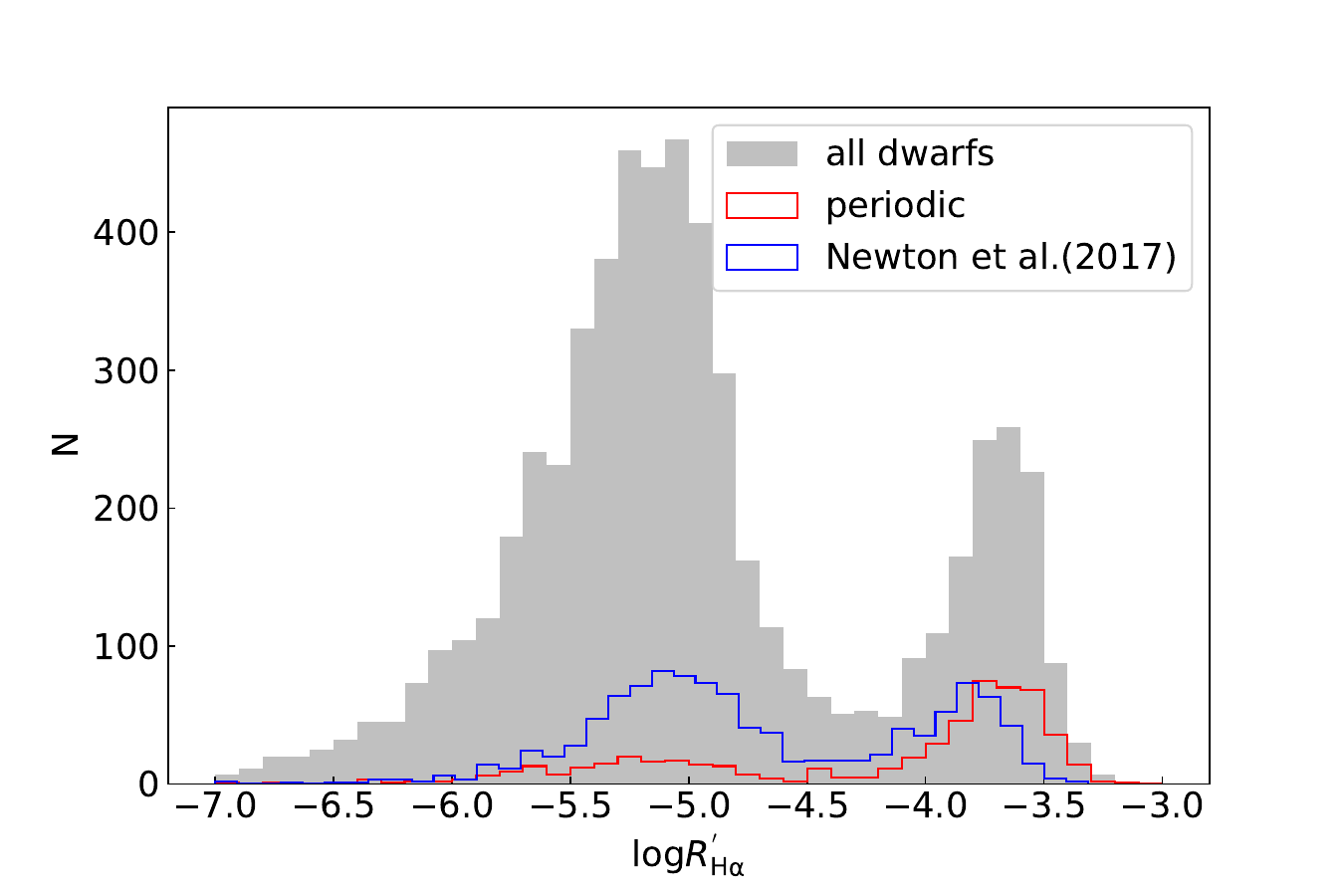}}
    \caption{Histogram of log$R^{\prime}_{\rm NUV}$ (Panel a), log$R^{\prime}_{\rm FUV}$ (Panel b), log$R^{\prime}_{\rm HK}$ (Panel c) and log$R^{\prime}_{\rm H\alpha}$ (Panel d). The gray filled bars represent the all dwarfs and the red line represent the periodic stars. The blue line in Panel c and d is the distribution of log$R^{\prime}_{\rm H\alpha}$ and log$R^{\prime}_{\rm HK}$ for M stars from \cite{2017ApJ...834...85N} and \cite{2018A&A...616A.108B}, respectively.}
    \label{R_distribution.fig}
\end{figure*}

Figure \ref{R_distribution.fig} shows the distributions of $R^{\prime}_{\rm NUV}$, $R^{\prime}_{\rm FUV}$, $R^{\prime}_{\rm HK}$, and $R^{\prime}_{\rm H\alpha}$.

There is a wide distribution value of log$R^{\prime}_{\rm NUV}$ ($\sim$ $-5$ to $-$2) and log$R^{\prime}_{\rm FUV}$ ($\sim$ $-5$ to $-$1).
Considering that the saturation regime for the UV bands are log$R^{\prime}_{\rm NUV}$ $\approx$ $-$3.5 and log$R^{\prime}_{\rm FUV}$ $\approx$ $-$3.9, the stars with quite high activity (log$R^{\prime}_{\rm UV}$ $\gtrsim$ $-$2.5) may be due to some contamination (see detailed discussion in Section \ref{compare.sec}).

There is a noticeable gap in the activity distribution of $\rm H\alpha$ and
Ca \scriptsize{\uppercase\expandafter{\romannumeral2}} \normalsize H$\&$K emission, similar to the Vaughn–Preston gap \citep{1980PASP...92..385V} discovered for F and G stars.
Possible explanations include sample incompleteness or a fast transition between the two populations.
The scenario of sample incompleteness can be ruled out since many previous studies \citep{2016ApJ...821...93N, 2020A&A...638A..20M, 2019ApJS..244...21S,2022ApJ...929...80B, 2023ApJS..264...12H}, using different samples and activity proxies, have also reported a similar double-peaked distribution for M stars.
The double peaks are quite consistent with the saturated region 
(log$R^{\prime}_{\rm{HK}} \approx -4$; log$R^{\prime}_{\rm H\alpha} \approx -3.5$)
and the unsaturated region
(log$R^{\prime}_{\rm{HK}} \approx -5$; log$R^{\prime}_{\rm H\alpha} \approx -5$).
Therefore, this gap is most likely due to a transition from the fast-rotating saturated population to the slow-rotating unsaturated population \citep{2016ApJ...821...93N,2016MNRAS.463.1844S,2022ApJ...929...80B}, which suggests a lack of stars with intermediate rotational periods and thus a discontinuous spin-down evolution \citep{2020A&A...638A..20M}.

\begin{figure*}[htp]
    \centering
    \subfigure[]{
    \label{p_ha.fig}
    \includegraphics[width=0.49\textwidth]{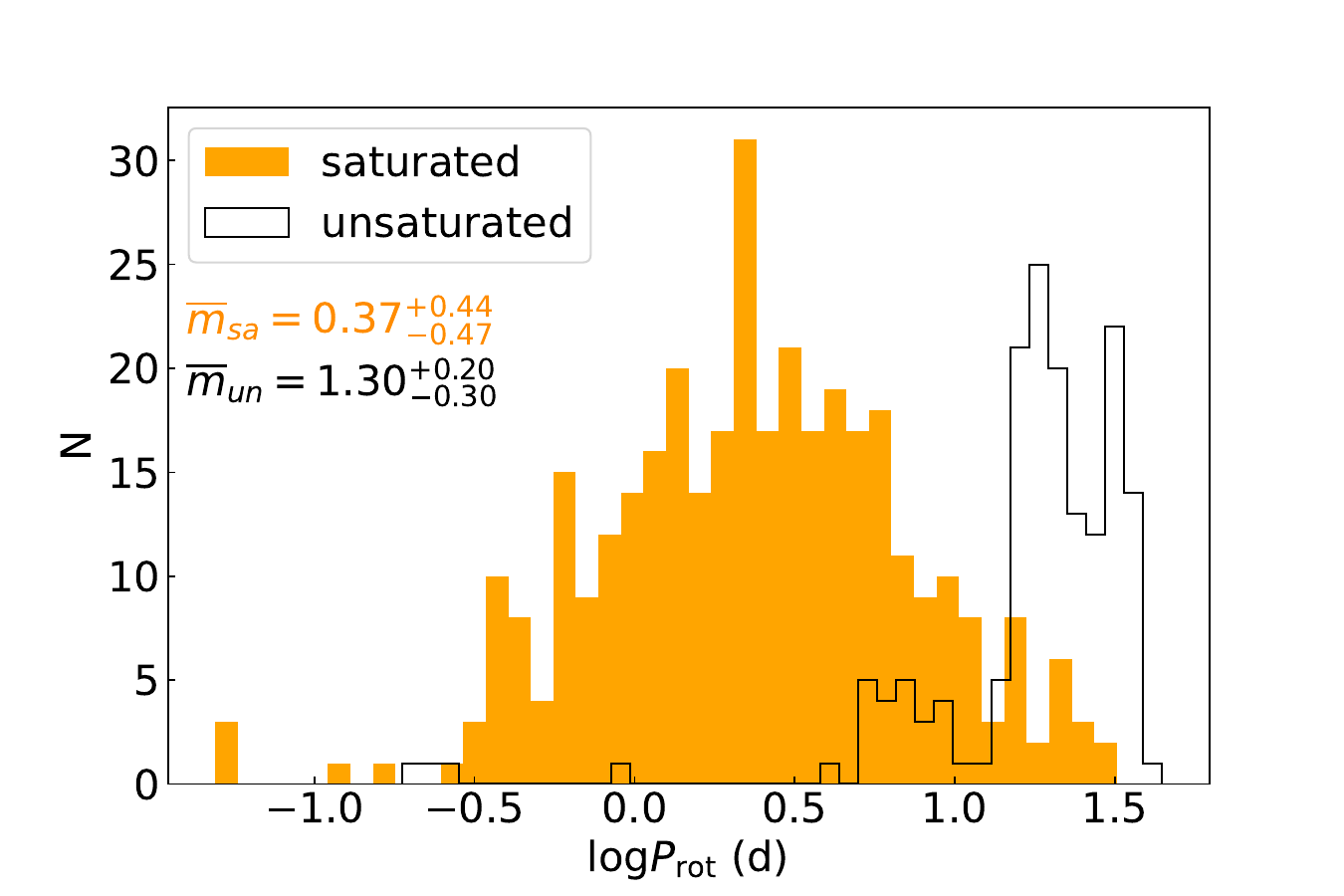}}
    \subfigure[]{
    \label{jz_ha.fig}
    \includegraphics[width=0.49\textwidth]{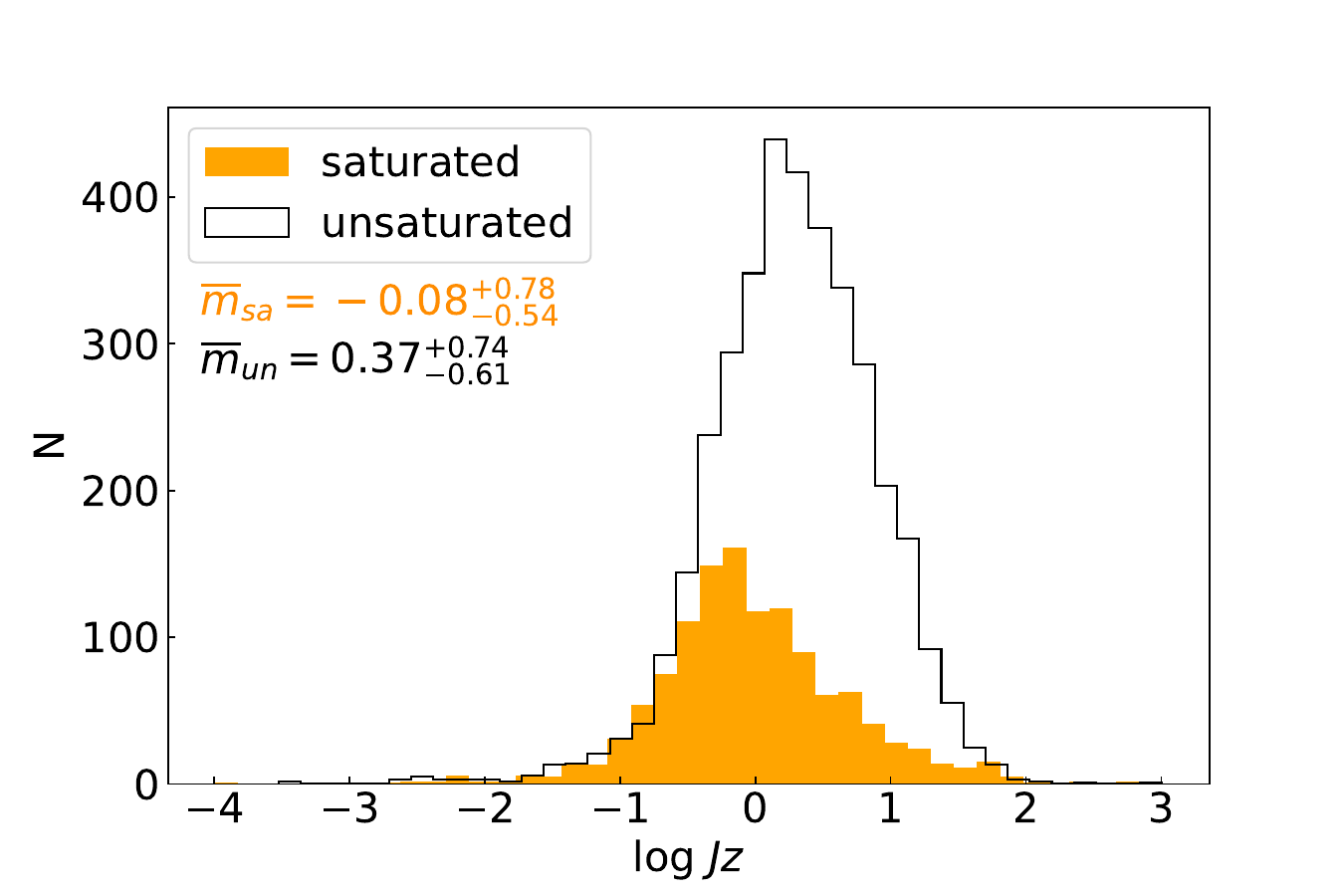}}
    \subfigure[]{
    \label{zmax_ha.fig}
    \includegraphics[width=0.49\textwidth]{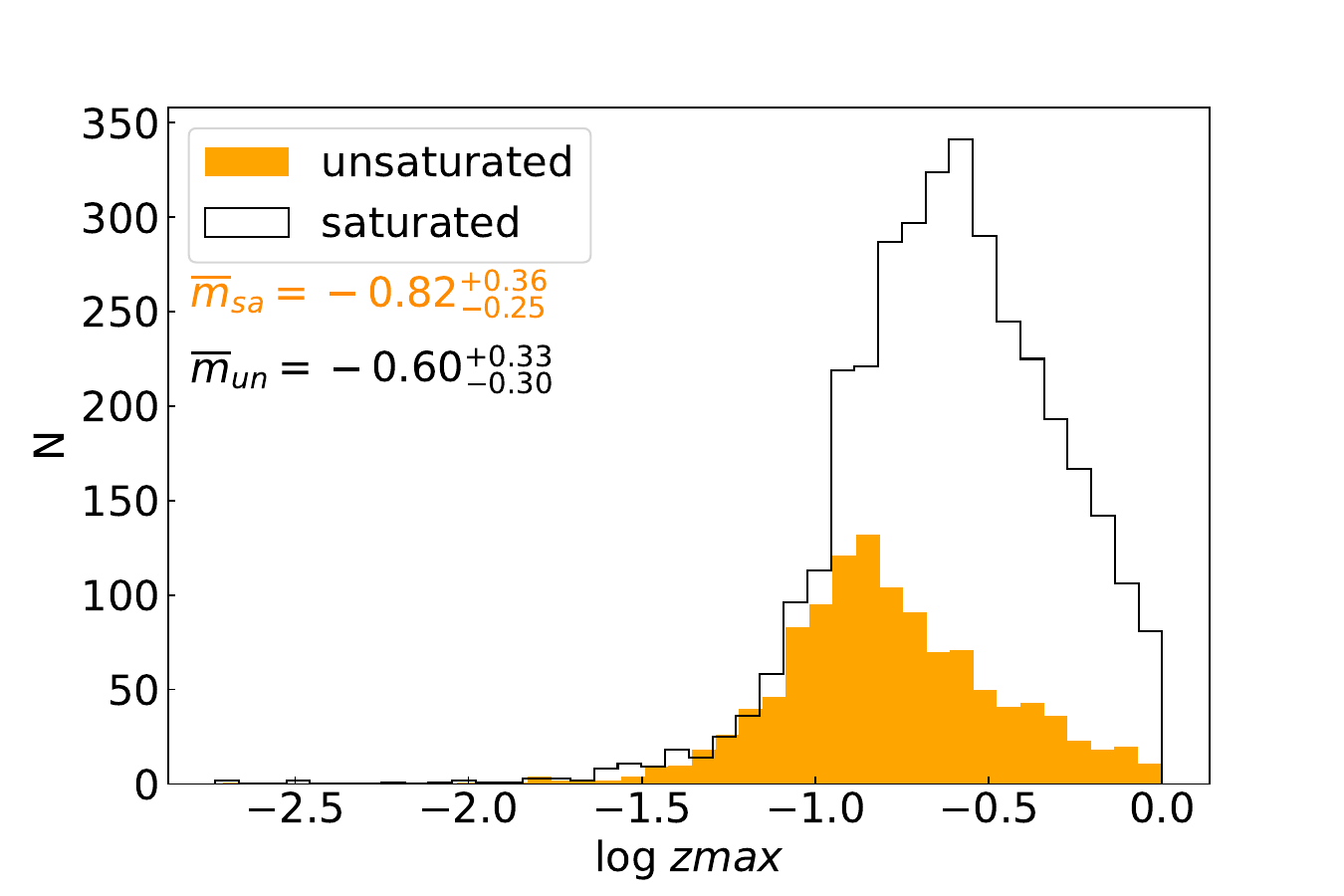}}
    \subfigure[]{
    \label{ecc_ha.fig}
    \includegraphics[width=0.49\textwidth]{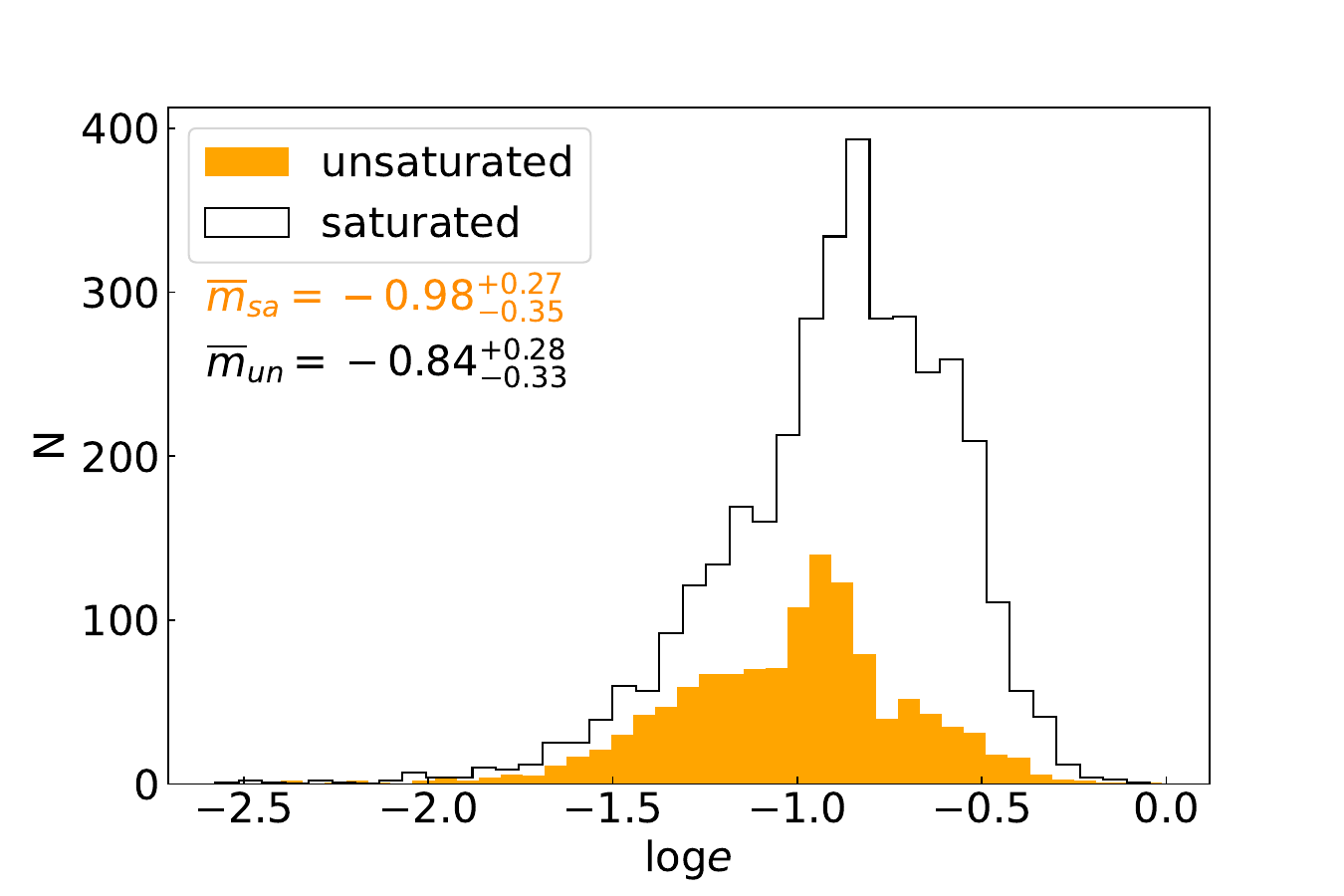}}
    \caption{Histogram of period (Panel a), $Jz$ (Panel b), $zmax$ (Panel c) and eccentricity (Panel d). The x-axis is the log scale. The orange filled bars represent the unsaturated stars (log$R^{\prime}_{\rm H\alpha} < -4.5$) and the black line represent the saturated stars (log$R^{\prime}_{\rm H\alpha} \geq -4.5$). The $\bar m_{un}$ and the $\bar m_{sa}$ represent the median value for the unsaturated sample and saturated sample, respectively.}
    \label{orbit.fig}
\end{figure*}

In order to investigate above scenario, we further examined the distribution of rotation periods $P_{\rm rot}$ and three galactic orbital parameters, including the vertical action $Jz$, maximum vertical height $zmax$, and eccentricity $e$ (Figure \ref{orbit.fig}).
These orbital parameters were measured with the $galpy$ package \citep{2015ApJS..216...29B}, under the St\"ackel approximation \citep{2012MNRAS.426.1324B} with the Milky Way potential MWPotential2014.
The clear difference in the rotation periods between the two populations (Figure \ref{orbit.fig}, Panel a) suggests that the scenario is plausible.
A rapid decay of the rotation period during a stage of stellar evolution leads to a significant weakening of stellar activity.
The three orbital parameters, which are approximate indicators of stellar age, indicate that the saturated population are generally (dynamically) younger than the unsaturated one, which is consistent with previous studies \citep{2011ApJ...727...56I, 2015ApJ...812....3W, 2016ApJ...821...93N}.
In addition, such a double-peaked distribution exists from M0 to M4 types, it is unlikely that the two populations are caused by distinct magnetic dynamos, such as the tachocline ($\alpha$--$\Omega$) and convective ($\alpha^2$) dynamos.

\begin{figure*}
    \centering
    \subfigure[]{
    \label{nuv_type.fig}
    \includegraphics[width=0.49\textwidth]{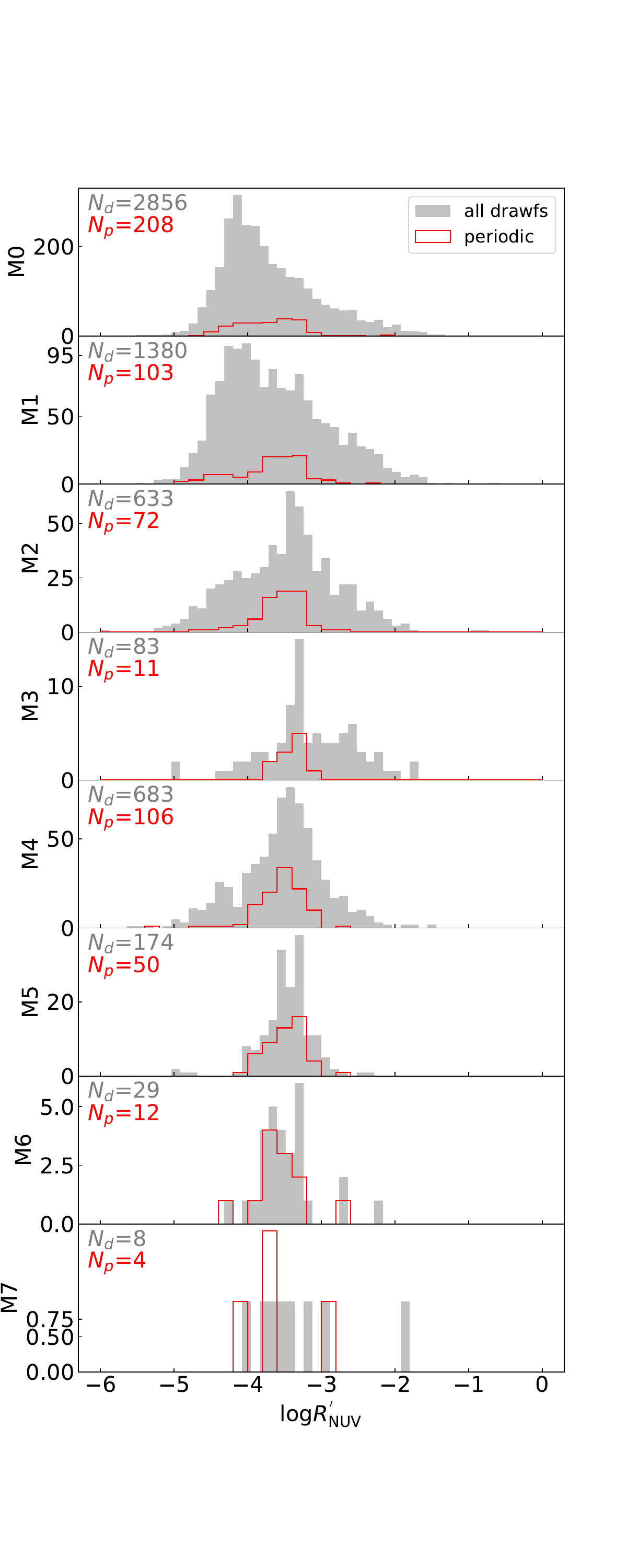}}
    \subfigure[]{
    \label{fuv_type.fig}
    \includegraphics[width=0.49\textwidth]{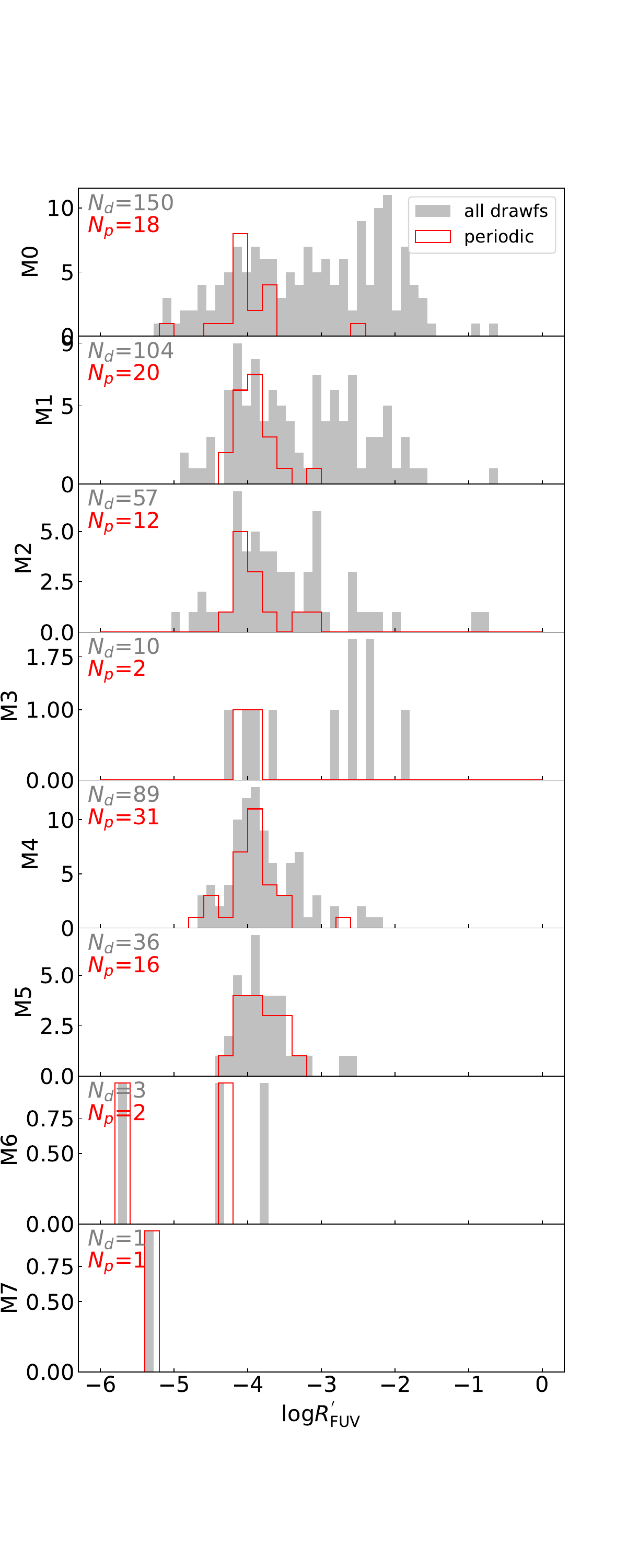}}
    \caption{Number distribution of log$R^{\prime}_{\rm NUV}$ (Panel a) and log$R^{\prime}_{\rm FUV}$ (Panel b) for stars from M0 to M7. The gray filled bars represent the main-sequence stars, the red line represent periodic sample.}
    \label{R_type.fig}
\end{figure*}

\begin{figure*}
    \centering
    \subfigure[]{
    \label{rhk_type.fig}
    \includegraphics[width=0.49\textwidth]{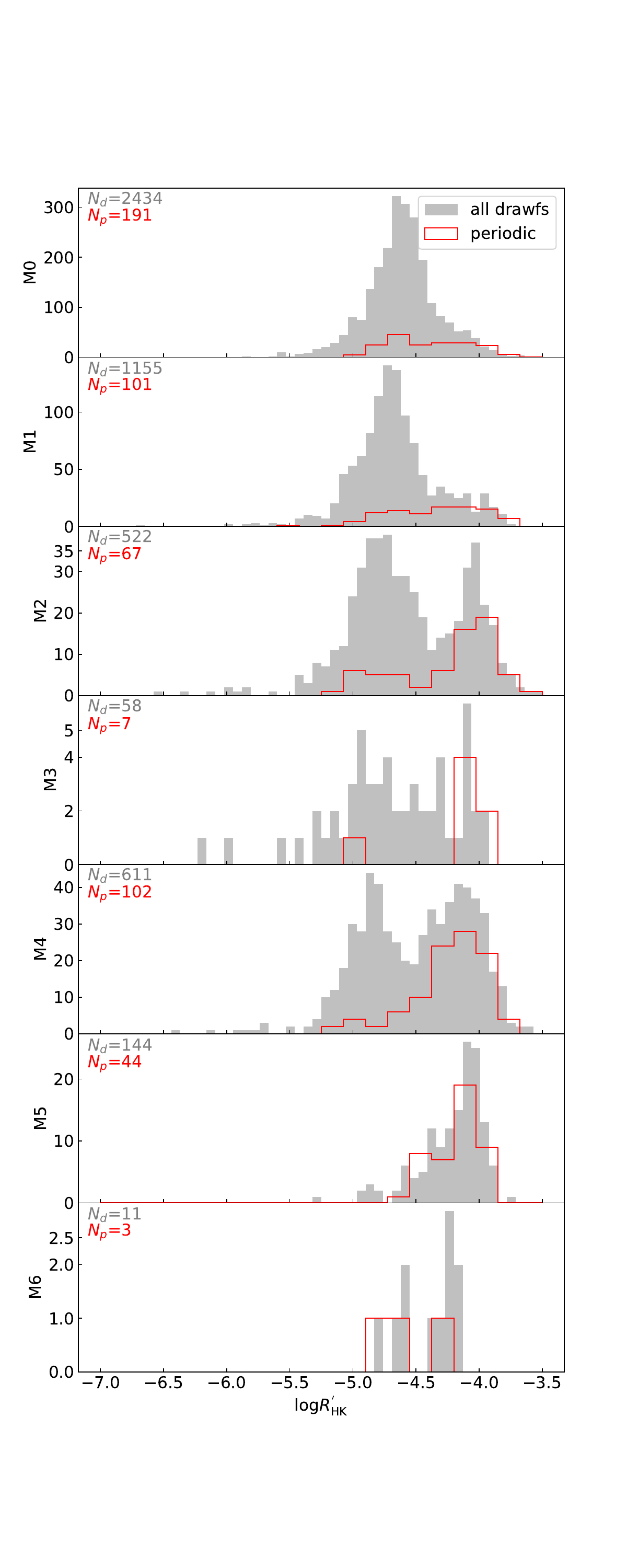}}
    \subfigure[]{
    \label{rha_type.fig}
    \includegraphics[width=0.49\textwidth]{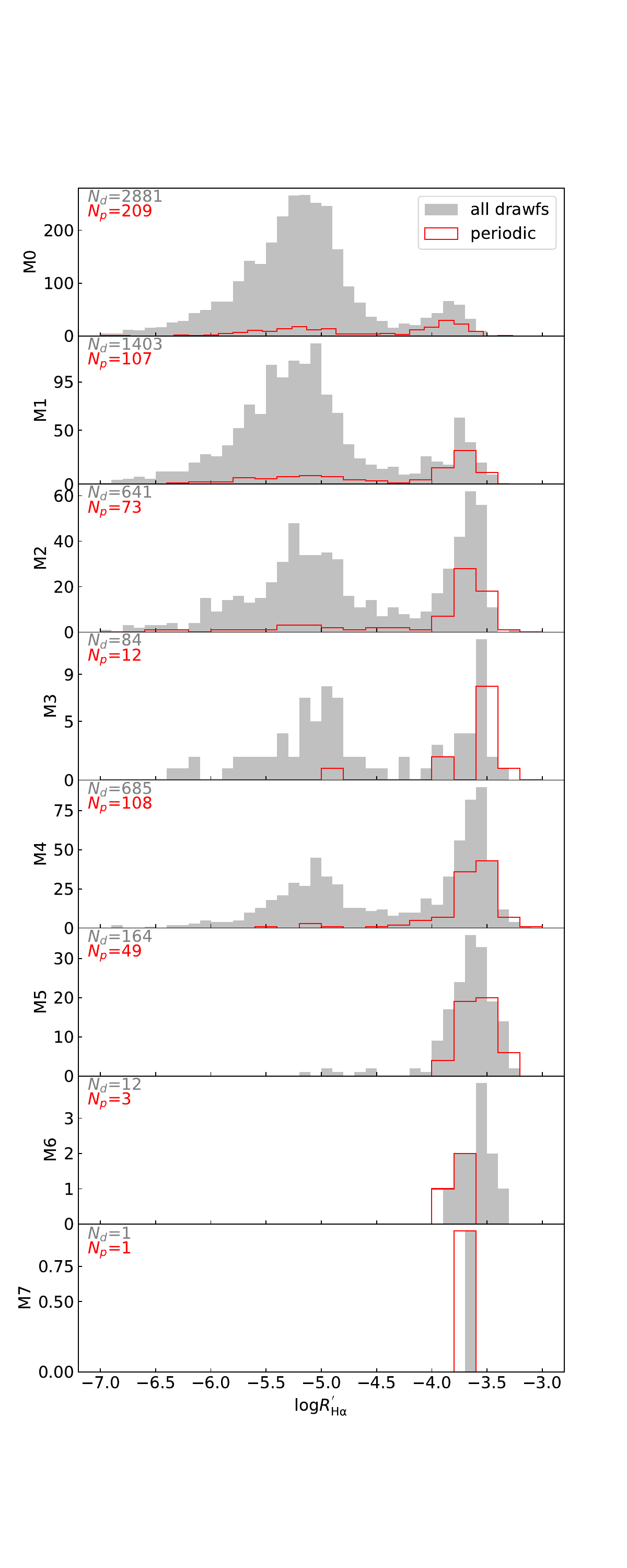}}
    \caption{Number distribution of log$R^{\prime}_{\rm HK}$ (Panel a) and log$R^{\prime}_{\rm H\alpha}$ (Panel b) for stars with different subtypes. The gray filled bars represent the main-sequence stars, and the red line represent the periodic sample.}
    \label{R_type_ha_hk.fig}
\end{figure*}

The distribution reveals a clear evolution from M0 to M6 types (Figure \ref{R_type.fig} and Figure \ref{R_type_ha_hk.fig}):
(1) for M0 and M1 types, most stars reside in the unsaturated region, and the double-peaked feature is weak for the Ca \scriptsize{\uppercase\expandafter{\romannumeral2}} \normalsize H$\&$K band;
(2) for M2 to M4 types, the portion of saturated and unsaturated population are approximately equal;
(3) for M5 and M6 types, the distribution evolves into a single peak located in the saturated region, with only a few stars ($\lesssim$10\%) remaining in the unsaturated region.
This can be explained by that different topology of magnetic fields results in  various stellar winds and angular momentum losses, leading to diverse spin-down rates \citep{2012ApJ...754L..26M, 2015ApJ...813...40G}.
The multipole magnetic field has a weaker magnetic breaking effect compared to a dipolar magnetic field, resulting in a more slowly decay of rotation in late-type M stars compared to early-type M stars \citep{2015ApJ...799L..23M}.
This also means both early-type (partly convective) stars and late-type (fully convective) stars operate rotation-dependent dynamos.
Furthermore, the smooth evolution of the distribution from M0 to M6 subtypes
imply a common dynamo for all the stars, in which the differential rotation and convection play more significant roles than the tachocline \citep{2016Natur.535..526W}.
It's worth exploring whether the abrupt variation of activity from mid-type (M2--M4) to late-type (M5--M6) stars (Figure \ref{R_type.fig}) is due to sample incompleteness (i.e., limited detection of very cool stars) or the operation of a distinct dynamo mechanism working for late-type stars.

\subsection{Relation between different activity indices}
\label{compare.sec}

\begin{figure*}[t]
   \center
   \includegraphics[width=0.98\textwidth]{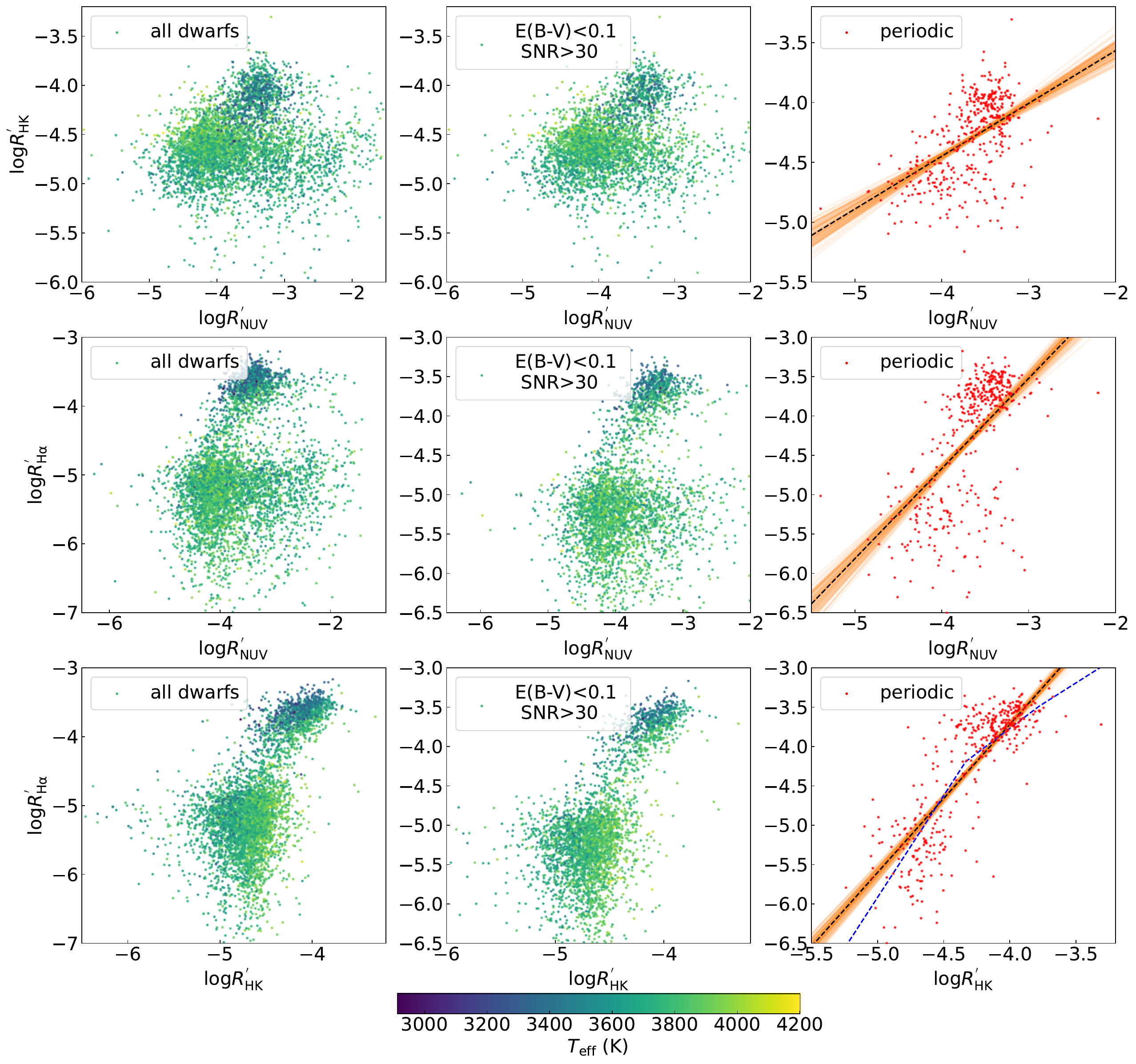}
   \caption{Top panels: Comparison of $R^{\prime}_{\rm NUV}$ and $R^{\prime}_{\rm{HK}}$. The samples from left to right panels are all dwarfs, targets with low extinction ($E(B-V) < 0.1$) and high S/N (S/N $>$ 30), targets with period measurements. The black dashed line is the fitting result.
   Middle panels: Comparison of $R^{\prime}_{\rm NUV}$ and $R^{\prime}_{\rm H\alpha}$. 
   Bottom panels: Comparison of $R^{\prime}_{\rm H\alpha}$ and $R^{\prime}_{\rm{HK}}$. The blue dashed line is the result from piecewise fitting.}
   \label{rhk_ha_nuv.fig}
\end{figure*}

Figure \ref{rhk_ha_nuv.fig} (top panels) shows the comparison between log$R^{\prime}_{\rm{HK}}$ and log$R^{\prime}_{\rm NUV}$. 
There is a positive but scattered relation, and the scatter is mainly caused by some objects with high levels of UV emission, which may be influenced by their surrounding environments.
The number of scatter points decreases for the sample with $E(B-V) < 0.1$ and $S/N > 30$ (middle panels).
The stars with rotational period measurements (right panel) show a clear correlation between the two indices.
A Markov Chain Monte Carlo (MCMC) fit was applied to these stars, and the result is
\begin{equation}
    {\rm log}R^{\prime}_{\rm{HK}} = 0.44^{+0.04}_{-0.04} \times {\rm log}R^{\prime}_{\rm NUV} - 2.68^{+0.14}_{-0.13}.
\end{equation}
Figure \ref{rhk_ha_nuv.fig} (middle panel) also shows a positive but diffuse relation between log$R^{\prime}_{\rm{H\alpha}}$ and log$R^{\prime}_{\rm NUV}$.
The MCMC fitting result of the relation using the periodic sample is
\begin{equation}
    \log R^{\prime}_{\rm{H\alpha}} = 1.15^{+0.08}_{-0.07} \times \log R^{\prime}_{\rm NUV} - 0.07^{+0.26}_{-0.27}.
\end{equation}

There is a clear and tight relation between  log$R^{\prime}_{\rm{H\alpha}}$ and  log$R^{\prime}_{\rm HK}$, shown in Figure \ref{rhk_ha_nuv.fig} (bottom panel).
A linear fitting using the periodic sample gives
\begin{equation}
    \log R^{\prime}_{\rm{H\alpha}} = 1.88^{+0.06}_{-0.06} \times \log R^{\prime}_{\rm HK} + 3.80^{+0.25}_{-0.26}.
\end{equation}
Furthermore, we also did a piecewise fitting between the log$R^{\prime}_{\rm{H\alpha}}$ and log$R^{\prime}_{\rm HK}$ because they have two populations, and the fitting result is
\begin{equation}
\label{hk_ha.eq}
 \log R^{\prime}_{\rm{H\alpha}} = 
\begin{cases}
 (2.59\pm 0.15) \times \log R^{\prime}_{\rm HK} + (7.01\pm 0.27), \\
 \quad {\rm if}\ \log R^{\prime}_{\rm HK} \le \rm -4.26\pm 0.06\\
 (1.06\pm 0.22) \times \log R^{\prime}_{\rm HK} + (0.53\pm 0.34), \\
 \quad {\rm if}\ \log R^{\prime}_{\rm HK} > \rm -4.26\pm 0.06.
 \end{cases}
\end{equation}

\begin{figure}[t]
    \centering
    \subfigure[]{
    \label{rgc_z.fig}
    \includegraphics[width=0.49\textwidth]{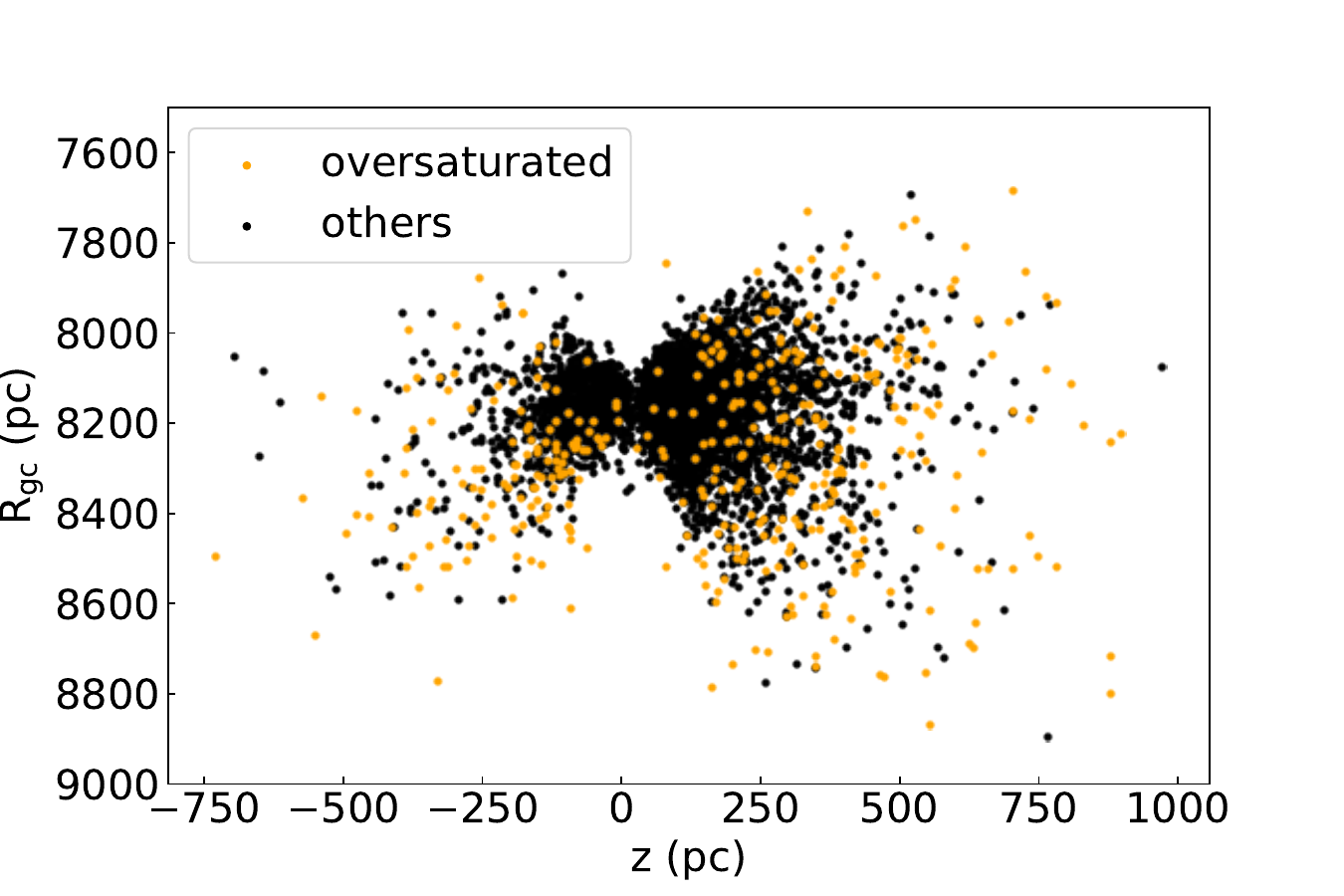}}
    \subfigure[]{
    \label{uvw.fig}
    \includegraphics[width=0.49\textwidth]{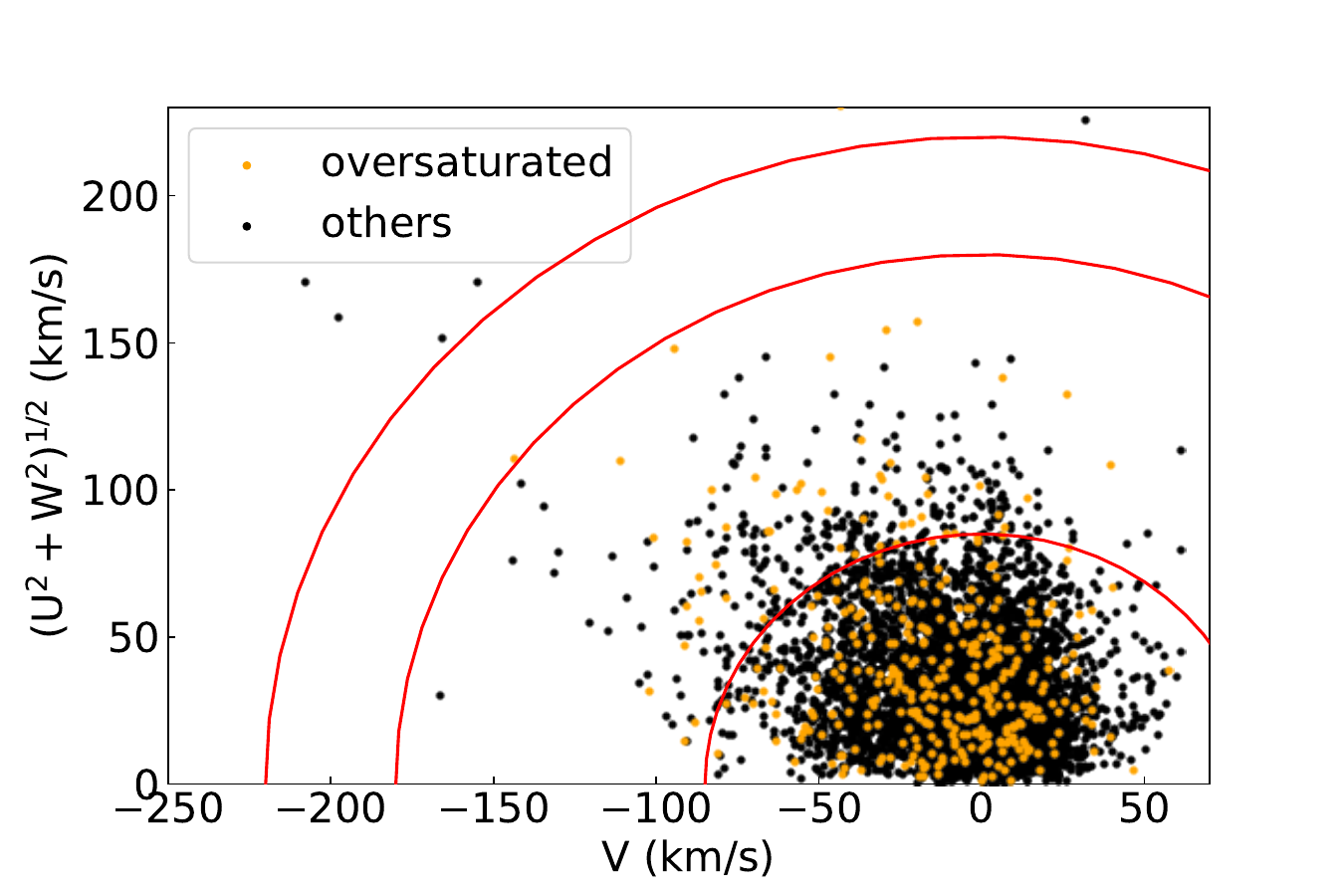}}
    \caption{Galactic parameters of the oversaturated population (yellow points) and other stars (black points).
    Panel (a): Galactocentric distance versus the distance to the Galactic plane.
    Panel (b): Toomre diagram. The three red lines represent constant values of the total Galactic velocity of 85, 180, and 220 km/s (from bottom to top), respectively.}
    \label{spatial_distribution.fig}
\end{figure}

We noticed a group about 400 stars (out of a total of 6629) show very high UV activity above the saturation value, characterized by log$R^{\prime}_{\rm UV}$ $\gtrsim$ $-$2.5 (hereafter ``oversaturated", also shown in Figure \ref{R_type.fig}), while exhibiting low levels of chromospheric activity.
Several studies \citep[e.g.,][]{2011ApJ...727....6S,2013MNRAS.431.2063S,2016ApJ...817....1J} have also identified groups of stars with high UV activity, but the underlying reason remains unknown.
Figure \ref{spatial_distribution.fig} shows the spatial distribution and Toomre diagram of our sample stars.
The majority of the population with high UV activity are located in the thin disk, like most of the sample stars.
Generally, these objects have a more scattered distribution.
When compared to stars with similar apparent magnitudes, these objects are notably positioned at larger distances from the galactic center and the galactic plane.

We considered several explanations including the presence of a white dwarf companion or a companion with a similar stellar type, contamination from surrounding environment, chance alignment with extragalactic sources, over-estimated extinction, or the possibility of a very young stellar population.
First, we did not find any UV spectral observations (e.g., from Hubble or IUE) for these sources.
By visually checking the LAMOST spectra of these stars, only 5\% of them show a possible excess in the blue band (Figure \ref{low_specs.fig}).
However, no wide Balmer absorption lines can be recognized.
This suggests that the scenario involving a white dwarf companion can not be the main reason.
Second, during the sample construction, we have employed a series of methods to identify and remove binaries. 
An examination of these stars on the HR diagram reveals that they do not fall within the binary belt, ruling out the scenario of a companion with a similar stellar type.
Third, we checked the DSS and PanSTARRS images of these sources and found that none of them are located in a nebula or star-forming region.
Forth, we cross-matched these sources with the GLADE+ galaxy catalog \citep{2018MNRAS.479.2374D} using match radii of 10/20/30 arcseconds, resulting in only 13/36/66 matches. This means the chance mismatch cannot account for the exceptionally active sample.
%
%
Fifth, almost all of these stars have $E(B-V)$ values less than 0.1, and their extinction uncertainties are small ($\lesssim 0.03$). Despite the large UV extinction coefficient, it is unlikely that extinction is the primary reason for the oversaturation of these stars.
Finally, although young stars (with an age of several million years) could exhibit very high UV activity \citep{2014AJ....148...64S, 2018AJ....155..122S}, most of these sources have large values of $zmax$ and $J_z$, suggesting that they are dynamically old.

\section{UV flare}
\label{flare.sec}

We further searched for flare events using the GALEX data, since they are also good indicators of stellar activity.
Firstly, we extracted the UV light curves of our targets using ``gPhoton", a software package that enables analysis of GALEX ultraviolet data at the photon level \citep{2016ApJ...833..292M}. 
There are a total of 10211 FUV and NUV observations for the 6629 sources.
Secondly, we tried to identify the flare events using the sigma clipping method, and
a flare event was identified when three consecutive data points exhibited a flux higher than 3$\sigma$ \citep{2019ApJS..241...29Y}.
We also found that the FLAIIL pipeline\footnote{https://github.com/parkus/flaiil} \citep{2018ApJ...867...71L}, which was developed for identifying flares in the FUV light curves from the MUSCLES data, is not suitable for many of our sources, due to that the exposure time is too short to establish the quiescent phase accurately.
Thirdly, we visually checked the light curve with possible flare events, and threw away the fake identifications and incomplete flares. 
Finally we derived 43 complete flare events for 35 stars, which are listed in Table \ref{flare.tab}. 

\begin{figure*}[t]
   \includegraphics[width=0.99\textwidth]{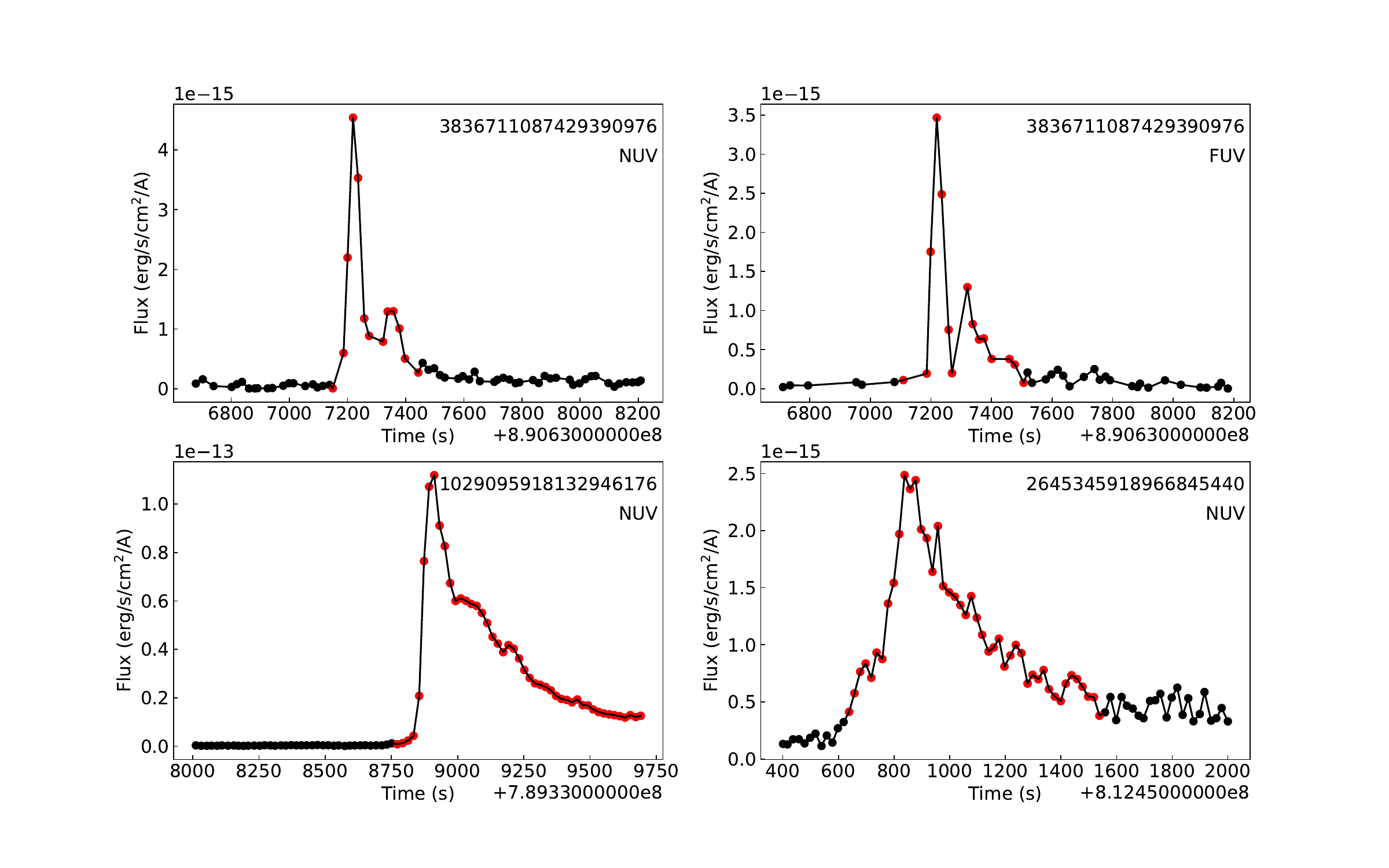}
   \caption{UV light curves of flare events. Upper panels: flare events observed simultaneously in the NUV and FUV bands for 3836711087429390976. Bottom panels: flares observed at different times in NUV band for 1029095918132946176 and 2645345918966845440. Black points are the quiescent state, red points are the flares.}
   \label{flare_lc.fig}
\end{figure*}

The durations of these flares range from $\approx$60 to $\approx$900 seconds, and the peak luminosities range from $10^{28}$ to $10^{31}$ erg/s. 
Most flares stars have subtypes ranging from M0 to M5.
Figure \ref{flare_lc.fig} displays some flares with typical signatures \citep{2007ApJS..173..673W}, including Type 1 flares (rapid rise and slow decay within 500 seconds) like 3836711087429390976, Type 2 flares (rapid rise and several peaks during the slow decay lasting longer than 500 seconds) like 1029095918132946176, and Type 3 flares (complicated shapes) like 2645345918966845440. 

\begin{table*}[]
\centering
\caption{All flare events in our sample.}
\begin{tabular}{ccccccc}
\hline\noalign{\smallskip}
Gaia id & Band & $\rm t_{\rm end}-t_{\rm start}$ & Equivalent duration & Peak luminosity & Flare energy & log$R^{\rm \prime}_{\rm flare}$\\
 & & (s) & (s) & log (erg/s) & log (erg) &  \\
\hline\noalign{\smallskip}
\multirow{2}*{3836711087429390976} & FUV & 160 & 4039 & 30.59 & 31.86 &  \multirow{2}*{-1.70}\\
 & NUV & 294 & 5215 & 30.70 & 32.58 &  \\
 \hline
\multirow{2}*{636520926431812096} & FUV & 364 & 1006 & 30.23 & 31.87 & \multirow{2}*{-2.16}\\
 & NUV & 362 & 1361 & 30.40 & 32.39 & \\
 \hline
\multirow{2}*{1545784091618733184} & FUV & 252 & 778 & 30.10 & 31.31 &\multirow{2}*{-2.60}\\
 & NUV & 282 & 1487 & 30.16 & 32.02 & \\
 \hline
\multirow{2}*{2645345918966845440} & NUV & 379 & 1957 & 28.93 & 30.96 & \multirow{2}*{-2.68}\\
 & NUV & 900 & 14162 & 29.23 & 31.82 & \\
 \hline
\multirow{2}*{2485820216434079360} & NUV & 165 & 32 & 29.22 & 30.12 & \multirow{2}*{-1.90}\\
 & NUV & 195 & 1815 & 30.37 & 32.01 & \\
 \hline
\multirow{3}*{1449481785146363648} & FUV & 476 & 747 & 30.23 & 31.59 & \multirow{3}*{-2.85}\\
 & NUV & 285 & 595 & 30.13 & 31.78 & \\
 & NUV & 350 & 2012 & 30.47 & 32.37 & \\
 \hline
\multirow{2}*{855090094138137088} & NUV & 77 & 79 & 28.75 & 30.01 & \multirow{2}*{-2.49}\\
 & NUV & 398 & 908 & 29.14 & 31.07 & \\
 \hline
2645381962332324992 & NUV & 283 & 879 & 30.93 & 32.70 & -1.75 \\
2698395946257300992 & FUV & 408 & 784 & 30.28 & 31.72 & -2.26 \\
1287410781917419264 & NUV & 180 & 723 & 30.22 & 32.00 & -2.34 \\
1282550012807310976 & FUV & 284 & 279 & 30.64 & 31.19 & -1.95 \\
3851331671501030784 & FUV & 196 & 255 & 29.83 & 31.21 & -2.62 \\
3836050766272513408 & NUV & 253 & 290 & 30.67 & 32.39 & -2.02 \\
2550158684794063744 & NUV & 318 & 36 & 29.70 & 30.71 & -3.09 \\
2575639321307107328 & FUV & 299 & 385 & 30.06 & 31.27 & -2.42 \\
3665691266433813248 & NUV & 275 & 191 & 28.99 & 30.60 & -3.12 \\
698725624975112192 & NUV & 457 & 981 & 31.10 & 33.03 & -1.81 \\
1449480616915236864 & NUV & 315 & 1168 & 30.75 & 32.21 & -1.80 \\
2561205585492641280 & FUV & 401 & 291 & 30.07 & 31.43 & -2.26 \\
2679196892688576384 & FUV & 101 & 241 & 29.12 & 30.48 & -2.99 \\
2644562860529295872 & NUV & 177 & 309 & 30.32 & 32.11 & -2.27 \\
4014961305479623424 & NUV & 239 & 403 & 29.45 & 31.34 & -2.65 \\
700016245467419136 & FUV & 745 & 150 & 29.34 & 30.91 & -3.03 \\
3648394677218755584 & NUV & 218 & 2135 & 29.06 & 31.05 & -2.52 \\
2815503108666238464 & NUV & 464 & 1039 & 28.81 & 30.89 & -2.86 \\
1479643038364982528 & NUV & 64 & 68 & 30.00 & 30.89 & -2.45 \\
684640777943741568 & NUV & 322 & 447 & 29.96 & 31.87 & -1.86 \\
666154478491325440 & NUV & 281 & 1647 & 29.23 & 31.17 & -2.72 \\
742553498486917888 & NUV & 481 & 776 & 29.45 & 31.63 & -2.25 \\
735728623654810624 & NUV & 479 & 702 & 29.49 & 31.49 & -1.77 \\
658601456379862784 & NUV & 321 & 321 & 29.07 & 30.71 & -2.30 \\
659265630123576448 & NUV & 186 & 397 & 29.51 & 31.12 & -2.19 \\
3266937637860051840 & NUV & 311 & 3094 & 29.22 & 30.86 & -1.64 \\
1314209831654576768 & FUV & 254 & 317 & 32.72 & 34.28 & -3.86 \\
\hline\noalign{\smallskip}
\end{tabular}

\label{flare.tab}
\end{table*}

We calculated the energy of all flare events as follows \citep{2023MNRAS.519.3564J,2023arXiv230617045R},
\begin{equation}
    E= 4 \pi d^{2} \delta \lambda \int_{t_{\rm start}}^{t_{\rm end}} (F(t)-F_{0}) dt,
\end{equation}
where $d$ is the distance from Gaia eDR3, $\delta \lambda$ is the effective bandwidths of FUV and NUV bands,  $t_{\rm start}$ and $t_{\rm end}$ are the times when the flare starts and ends, respectively. $F(t)$ is the measured flux at each time point during the flare, and $F_{0}$ is the estimated quiescent flux. 
We calculated the quiescent flux using an iterative method. First, the data point with the largest flux and adjacent points were removed. Then, two iterations were carried out to excluded the points with fluxes higher than 1$\sigma$. 
Finally, the median flux of the remaining points was calculated as the quiescent flux. 
The time corresponding to the first point with a flux larger than the quiescent flux marks the start of the flare ($t_{\rm start}$), and the time corresponding to the last point with a flux higher than the quiescent flux represents the end of the flare ($t_{\rm end}$).
The flare energy ranges from $10^{30}$ to $10^{34}$ erg, similar to the range of optical flares \citep[e.g.,][]{2017ApJ...849...36Y}. 
The flare activity index was calculated following $R^{\prime}_{\rm flare}=\frac{\Sigma E_{\rm flare}}{\int L_{\rm bol}dt}=\frac{\Sigma E_{\rm flare}}{L_{\rm bol} \times \Delta{t}}$ \citep{2017ApJ...849...36Y}.
Here $\Sigma E_{\rm flare}$ means the sum of the energy of all flare events for each source, and $\Delta{t}$ represents the total duration of these flares.

\section{Habitable Zone}
\label{hz.sec}

The habitability of exoplanets around M dwarfs is quite worthy of investigation due to the abundance of M-type stars in the Milky Way and the the extensive existence of exoplanets expected orbiting them \citep{2016PhR...663....1S}.

A planet was normally considered habitable when it resides in a region where suitable temperatures allow water to remain liquid on its surface.
This habitable zone is mainly determined by the properties of the host star (e.g., luminosity, effective temperature) and the distance between the planet and the star.
A continuously habitable zone \citep[CHZ:][]{1993Icar..101..108K, 2013ApJ...765..131K} was defined by establishing both an inner edge, calculated by the loss of water via photolysis and hydrogen escape, and an outer edge, determined by the maximum greenhouse due to $\rm CO_{2}$ clouds.

The habitable zone is also influenced by stellar magnetic activity, especially X-ray and UV emission and flares.
Thus, the habitability of planets surrounding M-type stars has been a topic of long-standing debate \citep{2016PhR...663....1S}.  
Although detailed effects of stellar activity on atmosphere of planets are not well understand, it has generally been believed that high activity (frequent flares and high levels of X-ray and UV activity) can be life-threatening, leading to atmospheric erosion \citep{2010A&A...511L...8S} and damaging biomolecules \citep{1973JThBi..39..195S}.
On the contrary, some experimental studies \citep[e.g.,][]{1977OrLi....8..259T, 2012NatCh...4..895R, 2015NatCh...7..301P} reported that UV emission can serve as a source of energy for prebiotic chemical synthesis, especially for ribonucleic acid, which is the building blocks for the emergence of life, while the low total UV emission of M stars may not support life processes like the chemical synthesis of complex molecules \citep{2017ApJ...843..110R,2018SciA....4.3302R}.
Taking these arguments into consideration, the concept of an ultraviolet habitable zone (UHZ) can be defined.

The overlapping region between CHZ and UHZ can be considered as the most favorable for habitability.
In this study, we first derived the CHZs for our sources using the method of \cite{2014ApJ...787L..29K}.
The inner edge of CHZ was calculated using the ``runaway greenhouse" limit (i.e., greenhouse effect caused by water), and the outer edge was determined using the ``maximum greenhouse" limit.
For these calculations we assumed a planet mass equivalent to that of Earth.
The CHZ was calculated following  
$d=(\frac{L/L_{\odot}}{S_{\rm eff}})^{0.5}$ AU \citep{2014ApJ...787L..29K},
where $S_{\rm eff}$ is the effective solar flux incident on the planet.
Then the UHZ was defined following \cite{2023MNRAS.522.1411S}:
the outer boundary of the UHZ was established with $f_{\rm UV} \geq$ 45 erg cm$^{-2}$ $s^{-1}$, and the inner boundary was set with $f_{\rm UV} \leq$ 1.04$\times10^4$ erg cm$^{-2}$ $s^{-1}$, twice the intensity of UV radiation that the Archean Earth received 3.8 billion years ago.

\begin{figure*}
    \centering
    \subfigure[]{
    \label{hz_bin.fig}
    \includegraphics[height=0.9\textwidth]{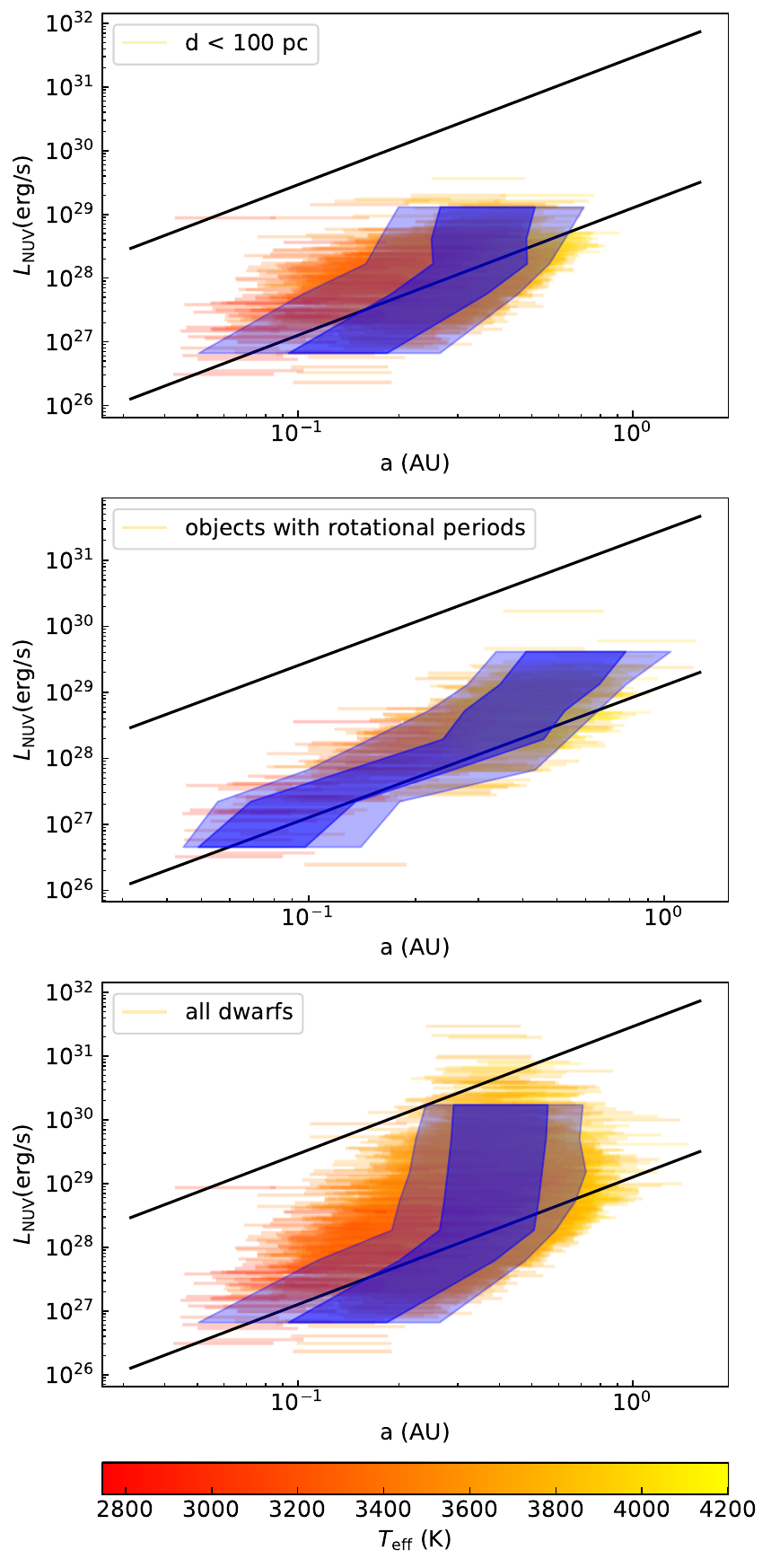}}
    \subfigure[]{
    \label{hz_d.fig}
    \includegraphics[height=0.9\textwidth]{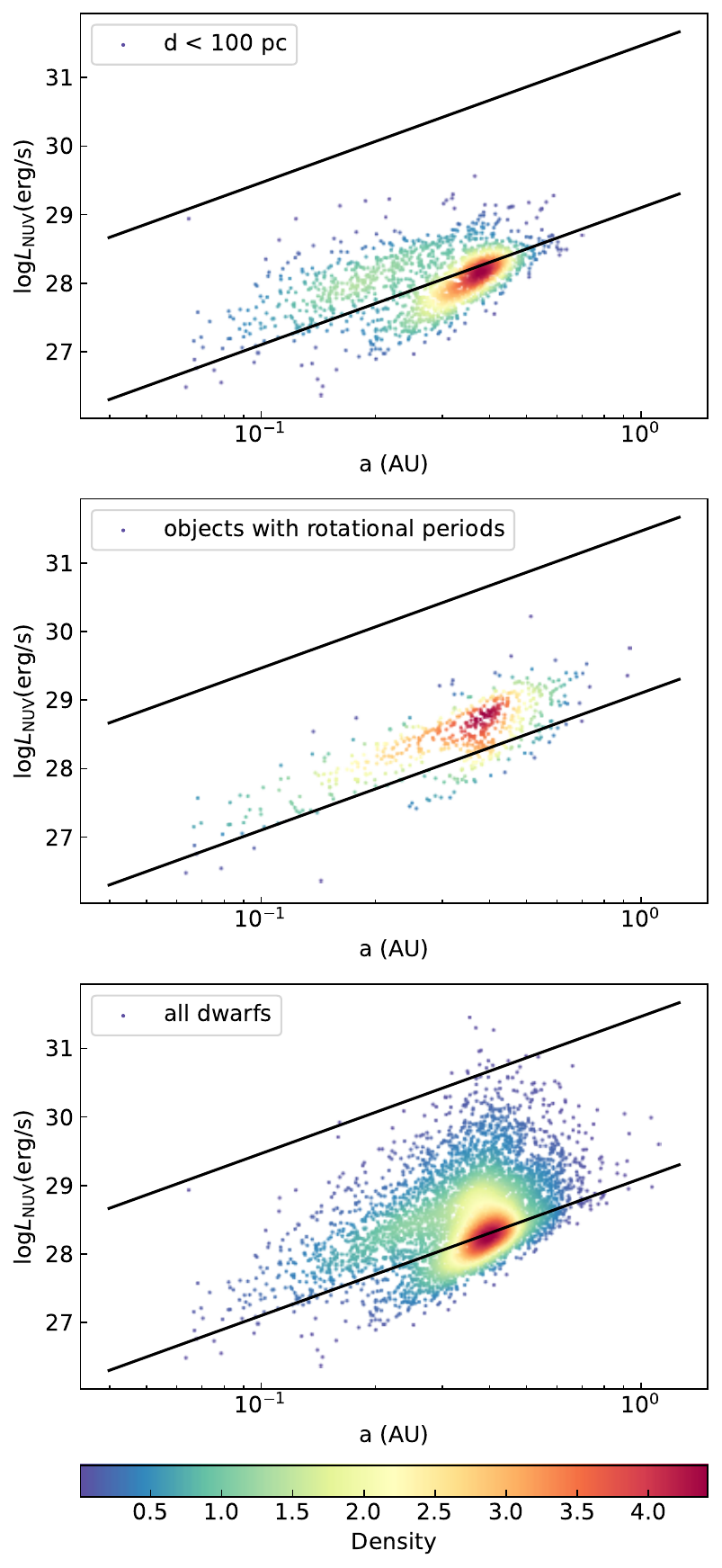}}
    \caption{The relation between NUV luminosity and the distance between star and planet. The horizontal lines in Figure \ref{hz_bin.fig} show the CHZ. The samples from top panel to bottom panel are the dwarfs within 100 pc, the stars with rotational periods, and all dwarfs. The color bar represent the different effective temperature of stars. The dark blue shaded area is the conservative CHZ, and the light blue shaded area is the more optimistic CHZ. The samples in Figure \ref{hz_d.fig} are as same as left, the color bar represent the density of stars. The black lines are the boundary of UHZ.}
    \label{hz.fig}
\end{figure*}

Figure \ref{hz.fig} shows the NUV luminosity $L_{\rm NUV}$ versus the star–planet distance $a$. 
The lower and upper black lines represent the inner and outer boundaries of the UHZ, respectively.
The orange lines represent the CHZs for each target.
Furthermore, we divided these stars into different bins based on their NUV luminosities.
For each bin, an conservative CHZ (dark blue shaded area) was calculated from the median values of the inner edges and the outer edges of the CHZs of each target.
Additionally, a more optimistic CHZ (light blue shaded area) was derived from the combination of the 16\% to 84\% values of the inner edges and the outer edges of the CHZs for each star.
The samples shown from the top panel to the bottom panel are M dwarfs within 100 pc, stars with rotational periods, and the total sample.
Table \ref{hz.tab} presents the probability of habitability for different samples.
Specifically, for nearby M stars within 100, 50, and 25 pc---a closer distance suggests a more complete sample, about 44\%, 42\%, and 41\% respectively, exhibit overlapping CHZ and UHZ regions.
Our results suggest for a significant number of M dwarfs, planets situated in the overlapping CHZ and UHZ regions are likely to be habitable.

\begin{figure}[t]
   \includegraphics[width=0.48\textwidth]{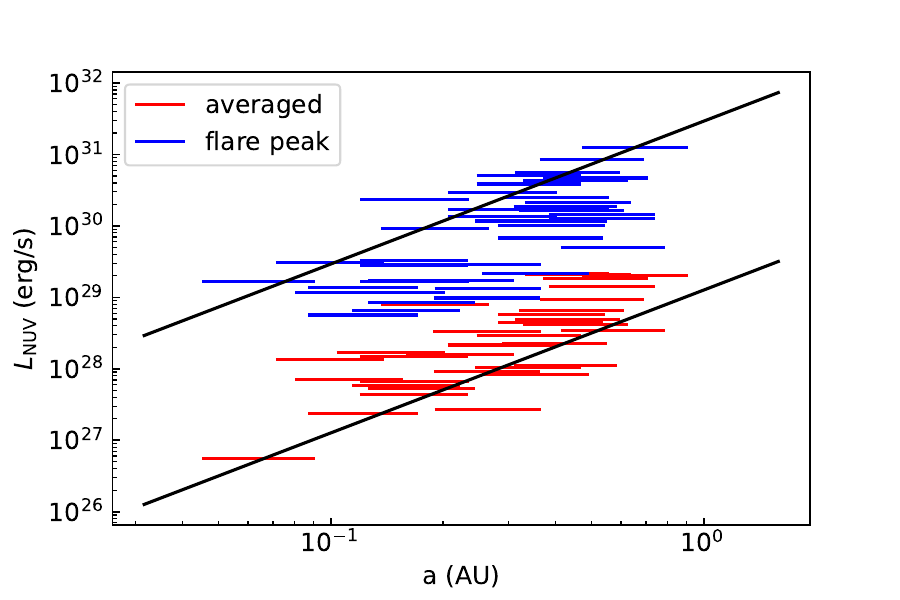}
   \caption{The relation between NUV luminosity and the star-planet distance of the sample with complete UV flares. The red lines are the normal luminosity of the stars, and the blue lines are the maximum luminosity of flare events.}
   \label{flare_hz.fig}
\end{figure}

We further examined the impact of stellar UV variability on its habitability.
We downloaded the data from multiple observations in the GALEX $VisitPhotoObjAll$ catalog using a match radius of 3$\arcsec$ via the CasJobs. 
This led to 3804 stars with multiple observations, with observation time spanning from 0.01 day to 9 years.
We replaced the NUV flux with the lowest and highest fluxes in multiple observations to access whether any of these stars would move outside the range of UHZ.
We found that around 0.31\% stars shifted outside the UHZ range. 
In addition, for the stars with UV flares, we calculated used the peak luminosities during the flare events.
Even we use the peak luminosity as the normal UV luminosity, most stars are still located in the UHZ region (Figure \ref{flare_hz.fig}).
In summary, UV variation had little impact on the probability of stellar habitability.

\begin{table}[t]
    \centering
    \caption{The number of stars and probability of habitability for different samples. N: The total number of stars. 
    P1: Probability of habitability by selecting star with overlapped CHZ and UHZ. 
    P2: Probability of habitability by selecting stars with the median value of CHZ falling within the UHZ. 
    P3: Probability of habitability by selecting stars with the CHZ completely within the UHZ.}
    \begin{tabular}{ccccc}
    \hline
         Sample & N & $\rm P_{1}$ & $\rm P_{2}$ & $\rm P_{3}$\\
         \hline
         < 25 pc & 27 & 67\% & 41\% & 26\%\\
         < 50 pc & 340 & 74\% & 42\% & 26\%\\
         < 100 pc & 1624 & 83\% & 44\% & 25\%\\
         Periodic & 567 & 97\% & 83\% & 64\%\\
         Total dwarfs & 5849 & 93\% & 68\% & 51\%\\
         \hline
    \end{tabular}
    \label{hz.tab}
\end{table}

\section{UV observation by CSST}
\label{discussion.sec}

CSST is a space-borne optical-UV telescope, which is scheduled to be launched around 2024 \citep{2023RAA....23g5009J}. 
It is designed with a primary mirror with a diameter of 2 meters.
CSST is equipped with seven photometric imaging bands and three spectroscopic bands, covering a wide range of wavelengths from the near-ultraviolet (NUV) to the near-infrared (NIR) \citep{2011SSPMA..41.1441Z}. 
CSST offers a large field of view (FOV) of approximately 1.1 deg$^{2}$ with a spatial resolution of $0.15^{"}$ for photometric imaging.
In the NUV band, CSST covers a wavelength range from 252 to 321 nm, with a  detection limit about 25 mag, much deeper than that of the GALEX telescope.
In order to explore the potential of studies on UV activity using CSST, we obtained the CSST NUV activity index from the SDSS $u$-band and GALEX NUV indices.

First, We cross-matched our sample with the SDSS DR16 catalog using TOPCAT, and found 5,770 targets with SDSS $u$-band magnitudes. 
The activity index of the SDSS $u$ band was calculated with steps similar to $R^{\prime}_{\rm NUV}$, following  
\begin{equation}
\label{activity_indexu.eq}
    R^{\prime}_{\rm u} = \frac{f_{\rm u,exc}}{f_{\rm bol}} = \frac{f_{\rm u,obs}\times(\frac{d}{R})^2-f_{\rm u,ph}}{f_{\rm bol}}
\end{equation}
The observed flux was estimated from $u$-band magnitude:
\begin{equation}
\label{f_u.eq}
    f_{\rm u}\ \rm{ (erg\ s^{-1}\ cm^{-2})} =  \rm 10^{-0.4\times(m_{u}+48.60-R_u \times E(B-V))} \times \delta \nu_{\rm u},
\end{equation}
where $\delta \nu_{\rm u}$ represents the frequency range corresponding to the effective bandwidth of the $u$ band, which was calculated as the range of wavelengths ($\approx$807.34 \AA) where the effective area falls to 10\% of its peak.
The $u$-band photospheric flux density was also derived using BT-Settl (AGSS2009) stellar spectral models.

\begin{figure}[t]
    \centering
    \includegraphics[width=0.49\textwidth]{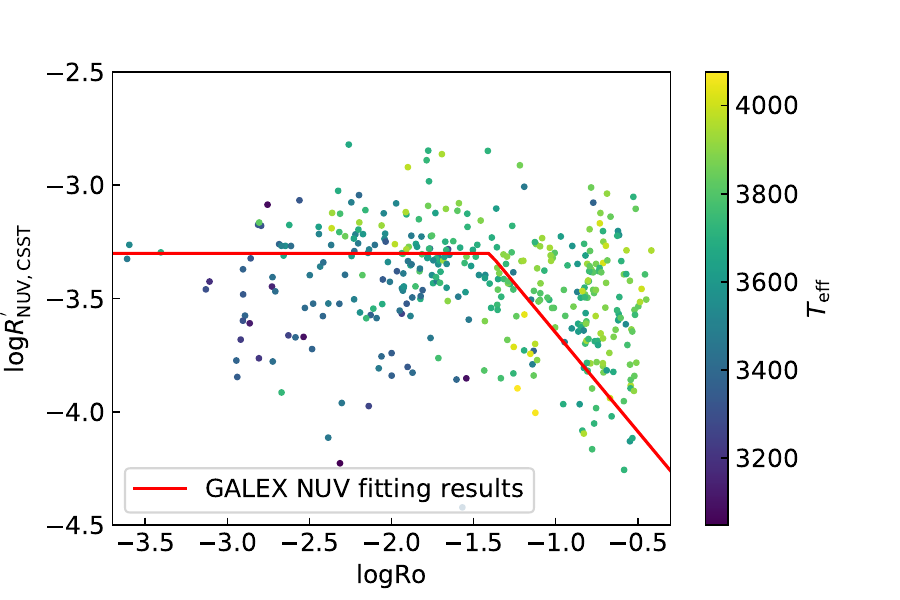}
    \caption{The activity-rotation relation of CSST NUV band. The red line is the fitting result from the GALEX NUV band.}
    \label{csst_ro.fig}
\end{figure}

Second, we obtained the CSST NUV activity index by interpolating between the SDSS $u$-band index and GALEX NUV index using their effective wavelengths
(i.e., 2877 \AA \quad for CSST NUV band, 2316 \AA \quad for GALEX NUV band, and 3608 \AA \quad for SDSS $u$ band).
Figure \ref{csst_ro.fig} shows the activity-rotation relation constructed from the CSST NUV band data.
The results are more diffuse compared to the GALEX NUV band index, which may be caused by the interpolation.

\begin{figure*}[t]
    \centering
    \includegraphics[width=0.8\textwidth]{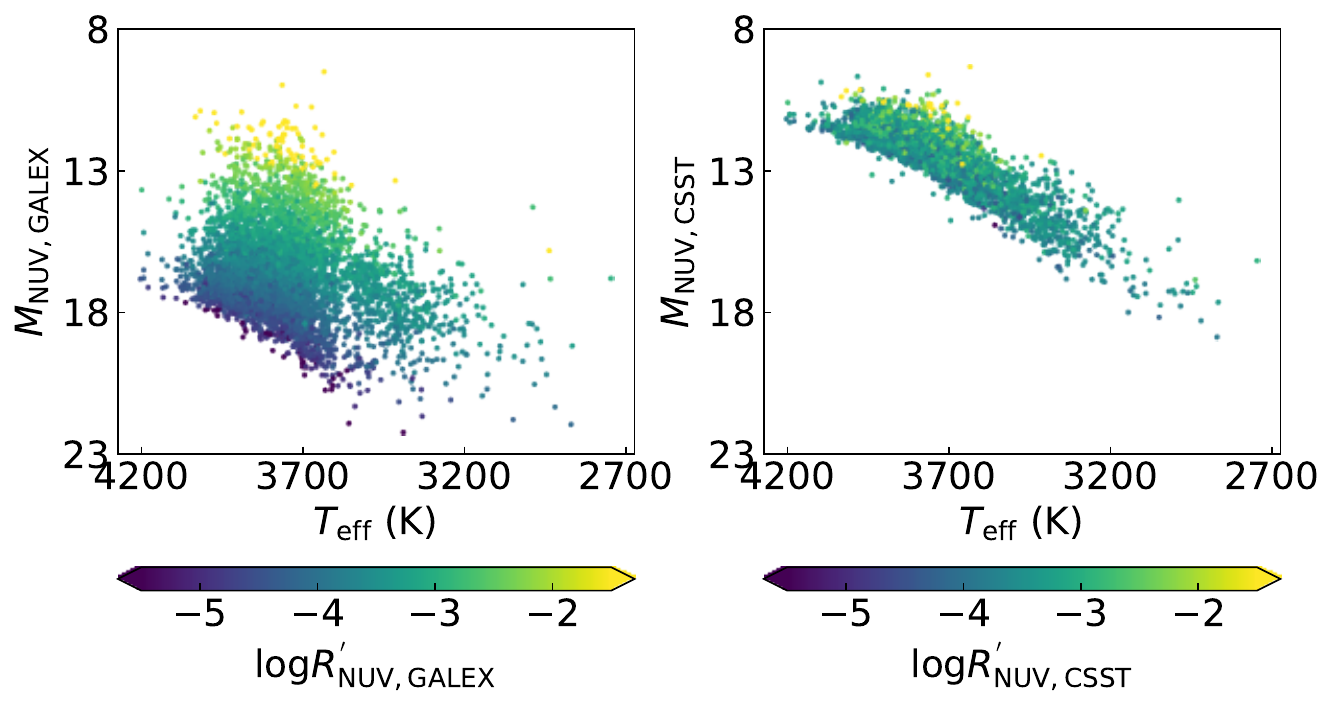}
    \caption{The magnitude vs. temperature diagram of GALEX NUV band (left panel) and CSST NUV band (right panel). The color bar of them are $R^{\prime}_{\rm NUV,GALEX}$ and $R^{\prime}_{\rm NUV,CSST}$, respectively.}
    \label{galex_csst_hr.fig}
\end{figure*}

Figure \ref{galex_csst_hr.fig} shows the magnitude versus temperature diagram for the GALEX NUV band (left panel) and CSST NUV band (right panel). The diffuse distribution of GALEX NUV magnitudes suggests dominance by various chromospheric emissions, while the tight distribution of CSST NUV magnitudes indicates emissions dominated by the photosphere.
In the temperature range of 3500--4000 K, the photospheric flux of the CSST NUV band accounts for about 60\%--80\% of the total flux. Accurate photometry and careful exclusion of the photospheric contribution are necessary for activity studies using the CSST NUV band.
Whatever, it is promising to investigate UV activity of numerous M stars (as far as 10--15 kpc) through CSST observations, particularly for faint stars that are below the detection limit of current telescopes.

\section{Summary}
\label{sum.sec}

By combining the LAMOST DR9 catalog and GALEX UV archive data, we studied chromospheric and UV activities of 6,629 M-type stars, including 5907 dwarfs, among which 582 ones have rotational period estimations, and 722 giants.

All the chromospheric and UV activity indices (i.e., $R^{\prime}_{\rm FUV}$, $R^{\prime}_{\rm NUV}$, $R^{\prime}_{\rm{HK}}$, and $R^{\prime}_{\rm{H\alpha}}$) clearly exhibit the saturated and unsaturated regions and the activity-rotation relation, in good agreement with previous studies.
Most cooler stars tend to occupy in the saturated region, while hotter ones are located in the unsaturated regime.
Both the FUV and NUV activity indices exhibit a wide single-peaked distribution. 
On the other hand, the {\rm H$\alpha$} and \rm {Ca \scriptsize{\uppercase\expandafter{\romannumeral2}} \normalsize H$\&$K} indices show a distinct double-peaked distribution.
The gap between these peaks is most likely due to a rapid transition from a fast-rotating saturated population to a slow-rotating unsaturated one, suggesting a lack of stars with intermediate rotational periods and thus a discontinuous spin-down evolution.
The clear difference in the rotational periods between the two populations further indicate that a rapid decay of the rotation period (during a stage of stellar evolution) leads to a significant weakening of stellar activity, especially for early-type M stars.
On the other hand, for the late-type M stars, the multipole magnetic field exhibits weak magnetic breaking, leading to a much slower rotation decay.
The smoothly varying distribution from M0 to M6 subtypes suggests a rotation-dependent dynamo for both early-type (partly convective) to late-type (fully convective) M stars.
In addition, the distributions of three galactic orbital parameters, including $Jz$, $zmax$, and $e$, indicate the saturated population are generally younger than the unsaturated one.

We examined the relationships between different activity proxies.
The FUV and NUV indices exhibit a tight relation described by ${\rm log} R^{\prime}_{\rm FUV} = (1.00\pm 0.02) \times {\rm log}R^{\prime}_{\rm NUV} - (0.37\pm 0.08)$.
The comparisons between the different proxies, especially $R^{\prime}_{\rm{H\alpha}}$ and $R^{\prime}_{\rm{HK}}$, reveal two subpopulations characterized by saturated and unsaturated.
A piecewise fitting between the $R^{\prime}_{\rm{H\alpha}}$ and $R^{\prime}_{\rm{HK}}$ is better than a linear fitting.
The relations between $R^{\prime}_{\rm NUV}$ and chromospheric indices are positive but scattered.
The scatter primarily arises from a group of $\approx$400 stars with oversaturated UV activity (log$R^{\prime}_{\rm NUV} > -$2.5) but unsaturated chromospheric activity. 
We considered several potential explanations for these stars, including the presence of a white dwarf companion, a companion with a similar stellar type, contamination from surrounding environment, chance alignment of extragalatic sources, over-estimated extinction, or the possibility of a very young stellar population.
However, none of these explanations could totally account for the characteristics of these stars.
Future UV spectral observations with the Space Telescope Imaging Spectrograph (STIS) or the Cosmic Origins Spectrograph (COS), both mounted on Hubble speace telescope, may help confirm whether there is a white dwarf companion \citep{2016MNRAS.463.2125P}.
If confirmed, this would represent a new method to detect white dwarfs in binaries lacking white dwarf features in the optical spectra---by selecting stars with abnormally high UV activity and normal chromospheric activity.

We searched for flare events in each GALEX exposure and identified 43 complete flare events of 35 stars. 
The durations of these flares vary from about 60 to 900 seconds; the peak luminosities range from $10^{28}$ to $10^{31}$ erg/s; the flare energy spans from $10^{30}$ to $10^{34}$ erg.
All of these properties are similar to those of optical flares \citep[e.g.,][]{2017ApJ...849...36Y}. 

The habitability of planets orbiting M-type stars is affected by stellar activity, especially ultraviolet activity. 
We calculated the CHZs and UHZs of each star, in order to investigate the proportion of habitable stars falling within the overlapping region of these two habitable zones.
We found 68\% M stars in the total sample and 44\%/42\%/41\% stars within 100/50/25 pc are potentially habitable.
The variation of UV luminosity due to random flux fluctuation or UV flare has a limited influence on the stellar habitability.

Finally, we calculated the stellar activity of SDSS $u$ band, and then obtained the stellar activity of CSST NUV band using interpolation. 
The typical (although somewhat scattered) activity-rotation relation of CSST NUV band suggests the potential for conducting UV activity studies through CSST observations in the future, especially for faint stars that fall below the detection limit of current telescopes.

\section*{acknowledgements}

We thank the anonymous referee for helpful comments and suggestions that have significantly improved the paper. 
We thank Dr. Riccardo Spinelli for very help discussions on stellar habitability.
The Guoshoujing Telescope (the Large Sky Area Multi-Object Fiber Spectroscopic Telescope LAMOST) is a National Major Scientific Project built by the Chinese Academy of Sciences. Funding for the project has been provided by the National Development and Reform Commission. LAMOST is operated and managed by the National Astronomical Observatories, Chinese Academy of Sciences.
Some of the data presented in this paper were obtained from the Mikulski Archive for Space Telescopes (MAST).
This work presents results from the European Space Agency (ESA) space mission {\it Gaia}. {\it Gaia} data are being processed by the {\it Gaia} Data Processing and Analysis Consortium (DPAC). Funding for the DPAC is provided by national institutions, in particular the institutions participating in the {\it Gaia} MultiLateral Agreement (MLA). The {\it Gaia} mission website is https://www.cosmos.esa.int/gaia. The {\it Gaia} archive website is https://archives.esac.esa.int/gaia. We acknowledge use of the VizieR catalog access tool, operated at CDS, Strasbourg, France, and of Astropy, a community-developed core Python package for Astronomy (Astropy Collaboration, 2013). 
This work was supported by National Key Research and Development Program of China (NKRDPC) under grant Nos. 2019YFA0405000 and 2019YFA0405504, Science Research Grants from the China Manned Space Project with No. CMS-CSST-2021-A08, Strategic Priority Program of the Chinese Academy of Sciences undergrant No. XDB4100000, and National  Natural Science Foundation of China (NSFC) under grant Nos. 11988101/11933004/11833002/12090042/12273057. S.W. acknowledges support from the Youth Innovation Promotion Association of the CAS (IDs 2019057).

\bibliographystyle{aasjournal}
\bibliography{bibtex.bib}{}

\begin{appendix}

\section{LAMOST spectra of stars with super high UV emission}

Here we shows some examples of the LAMOST low-resolution spectra for stars with extremely high UV emission (log$R^{\prime}_{\rm NUV}>-2.5$).

\renewcommand\thefigure{\Alph{section}\arabic{figure}}
\setcounter{figure}{0}
\begin{figure*}[h]
    \center
    \includegraphics[width=0.95\textwidth]{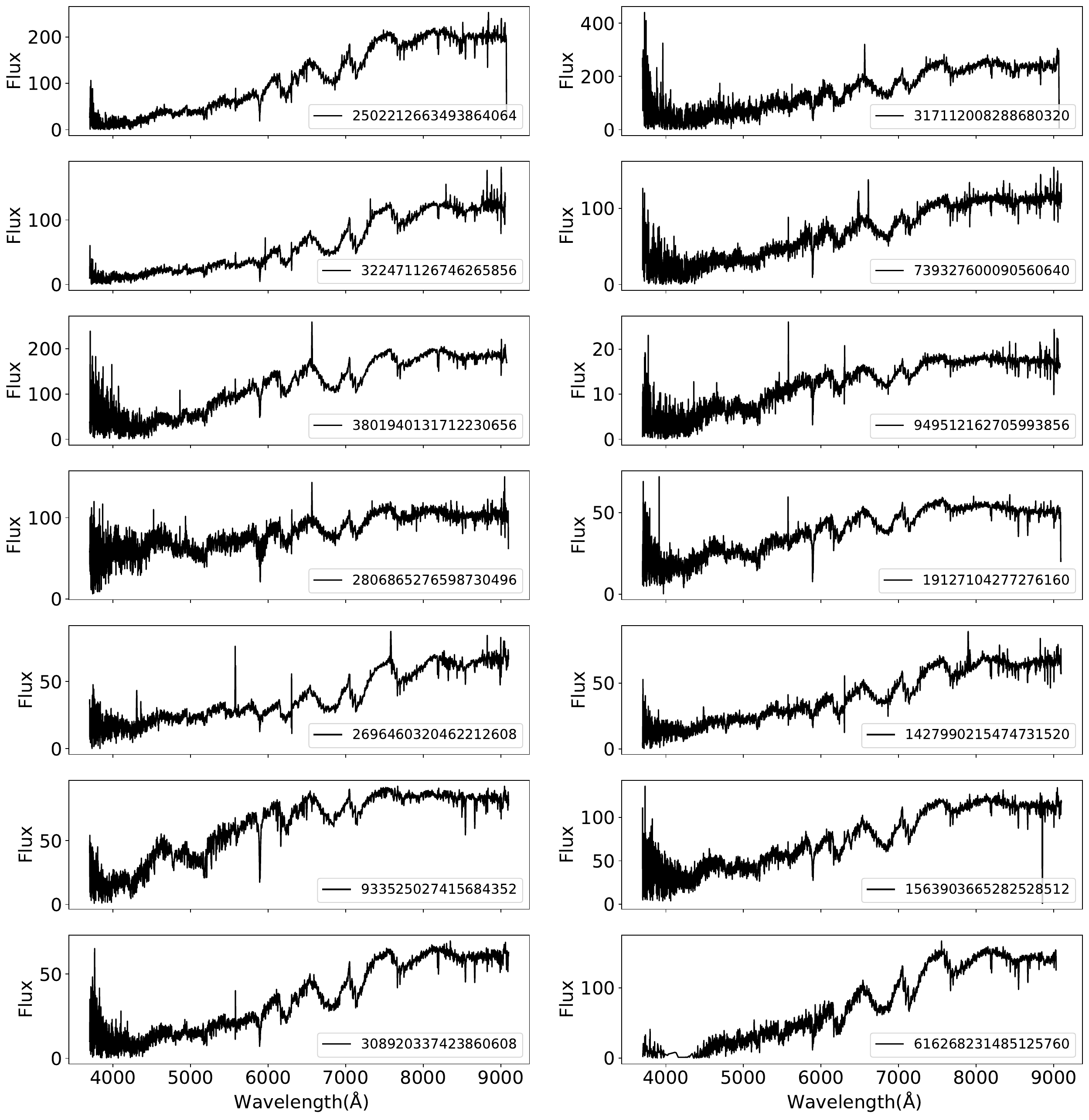}
    \caption{Examples of the LAMOST low-resolution spectra for stars with log$R^{\prime}_{\rm NUV}>-2.5$.}
    \label{low_specs.fig}
\end{figure*}

\section{Activities of giants}
\label{giants.sec}

There are 722 giants in the total sample of 6629 stars. 
We considered a star as a giant if its surface gravity (log$g$) value is smaller than 3.5.
The atmospheric parameters of M giants observed by LAMOST, especially log$g$ and [Fe/H], have a significant systematic offset with other surveys, such as APOGEE.
Therefore, we conducted a cross-match with APOGEE DR17\footnote{\url{https://www.sdss4.org/dr17/irspec/spectro_data/}} to obtain atmospheric parameters for the common sources. For the remaining giants, we calculated the [Fe/H] using the correlation between the $W1-W2$ color index and metallicity \citep{2016ApJ...823...59L}.
We calculated the activity indices of giants using the same method for dwarfs (Table \ref{all_pars.tab} and \ref{all_results.tab}).
Giants exhibit much lower activity level (the median of log$R^{\prime}_{\rm NUV}$ is about $-$5) compared to dwarfs (the median of log$R^{\prime}_{\rm NUV}$ is about $-$3.5).

\end{appendix}

\end{document}